\documentclass[]{jfm}

\usepackage[format = plain, singlelinecheck = no, justification = justified]{caption}
\usepackage{graphicx}
\usepackage{newtxtext}
\usepackage{newtxmath}
\usepackage{natbib}
\usepackage{hyperref}
\hypersetup{
    colorlinks = true,
    urlcolor   = blue,
    citecolor  = black,
}

\newcommand{\RomanNumeralCaps}[1]
\linenumbers

\title{Wall-modeled large-eddy simulation of three-dimensional turbulent boundary layer in a bent square duct}

\author{Xiaohan Hu\aff{1}
  Imran Hayat\aff{1}
 \and George Ilhwan Park\aff{1}
 \corresp{\email{gipark@seas.upenn.edu}},
 }

\affiliation{\aff{1}Department of Mechanical Engineering and Applied Mechanics, University of Pennsylvania, Philadelphia, PA 19104, USA\\}

\begin{document}
\maketitle

\begin{abstract}
We conduct wall-modeled LES (WMLES) of a pressure-driven three-dimensional turbulent boundary layer (3DTBL) developing on the floor of a bent square duct to investigate the predictive capability of three widely used wall models, namely, a simple equilibrium stress model, an integral nonequilibrium model, and a PDE nonequilibrium model. The numerical results are compared with the experiment of Schwarz \& Bradshaw (\textit{J. Fluid Mech.} (1994), vol. 272, pp. 183–210). While the wall-stress magnitudes predicted by the three wall models are comparable, the PDE nonequilibrium wall model produces a substantially more accurate prediction of the wall-stress direction, followed by the integral nonequilibrium wall model. 
The wall-stress direction from the wall models is shown to have separable contributions from the equilibrium stress part and the integrated nonequilibrium effects, where how the latter is modeled differs among the wall models. The triangular plot of the wall-model solution reveals different capabilities of the wall models in representing variation of flow direction along the wall-normal direction. On the contrary, the outer LES solution is unaffected by  the type of wall model used, resulting in nearly identical predictions of the mean and turbulent statistics in the outer region for all the wall models. This is explained by the vorticity dynamics and the inviscid skewing mechanism of generating the mean three-dimensionality. Finally, the LES solution in the outer layer is used to study the anisotropy of turbulence. In contrast to the canonical 2D wall turbulence, the Reynolds stress anisotropy exhibit strong non-monotonic behavior with increasing wall distance.
\end{abstract}

\begin{keywords}
Authors should not enter keywords on the manuscript, as these must be chosen by the author during the online submission process and will then be added during the typesetting process (see \href{https://www.cambridge.org/core/journals/journal-of-fluid-mechanics/information/list-of-keywords}{Keyword PDF} for the full list).  Other classifications will be added at the same time.
\end{keywords}

{\bf MSC Codes }  {\it(Optional)} Please enter your MSC Codes here

\section{Introduction}
\label{sec:intro}
The capability to predict high Reynolds number turbulent flows is essential for many natural and engineering flows such as external aerodynamics of wind turbines and aircraft wings, flow over the hull of marine vehicles, atmospheric boundary-layer flow over complex landscapes and cityscapes, to name a few. However, due to extreme disparity of scales present in high Reynolds number wall-bounded turbulent flows, any attempt to simulate these flows directly on a computational grid without resorting to modeling of some sort results in prohibitively large computational cost. To resolve all the scales of the near-wall turbulent motions using direct numerical simulation (DNS), the required number of grid points scales as $O(Re^{37/14})$, where $Re$ is the characteristic Reynolds number. Wall-resolved large-eddy simulation (WRLES), which resolves only the large (stress-carrying) eddies, reduces the grid point requirement to $O(Re^{13/7})$ \citep{Choi2012}. However, this level of computational cost is still unaffordable when the Reynolds number is high. As a cost-effective alternative to the above two approaches, wall-modeled large-eddy simulation (WMLES) resolves only the energetic eddies in the outer portion of the boundary layer, while the momentum transport  within the unresolved near-wall region is accounted for by augmenting the wall-flux through a wall model. Thus, the no-slip condition at the wall is replaced by the Neumann boundary condition supplied by the wall model in the form of the wall shear stress. Note that the wall shear stress is computed by the wall model without ever resolving the near-wall scales. Therefore, the grid point requirement for WMLES reduces to $O(Re)$ \citep{Choi2012}, making the simulations of high Reynolds number flows feasible.

To date, several wall models have been proposed, most of which are based on some form of the law-of-the-wall or solving a set of simplified or full Reynolds-averaged Navier-Stokes (RANS) equations. \cite{Deardorff} and \cite{Schumann} were the first to recognize the need for wall modeling to perform LES of high Reynolds number plane channels and annuli  to overcome lack of computing resources in the 1970s. \cite{grotzbach1987direct} later improved the model by removing the necessity of a priori knowledge of  the mean wall shear stress. 
The geometry of the near-wall eddies was incorporated in the work of \cite{shift}  
to account for the inclination of the vortical structures in the streamwise/wall-normal plane. 
\citet{Wang2002} proposed an ordinary differential equations (ODE) based wall model derived from the equilibrium assumption \citep{Degraaff2000}, which later on was extended to compressible flows \citep{Kawai2012,Bodart2011}. 
The ODE equilibrium wall model excludes the nonequilibrium effects such as pressure gradient, and considers the wall-normal diffusion only. 
Nonequilibrium wall models based on full 3D RANS equations were investigated by \citet{Balaras1996}, \citet{Wang2002}, \citet{Cabot2000}, \citet{Kawai2013}, and \citet{Park2014,Park2016}.
\color{black}
\citet{Yang2015} introduced the integral nonequilibrium wall model based on the integrated boundary layer equations and assumed velocity profiles, which can be considered as a compromise between the aforementioned two classes of wall models. Several efforts have also been directed toward formulating wall models which are not based on RANS. \citet{Bose2014} and \cite{bae2019dynamic} proposed a differential filter-based wall model which introduced a slip-velocity applied in the form of Robin boundary condition at the wall. \citet{Chung2009} proposed a virtual wall model with a slip velocity boundary condition specified on the lifted virtual wall. \citet{Gao2019} extended this virtual wall model in a generalized curvilinear coordinate. Advances on WMLES were reviewed by \citet{Piomelli2002}, and more recently by \citet{Larsson2016} and \citet{Park2018}. 
With the development of novel wall models and increase in the computing capacity, 
WMLES is becoming an indispensable tool for predictive but affordable 
scale-resolving simulation of practical engineering flows at high Reynolds numbers. Recent applications to external aerodynamics applications include simulation of a wing-body junction flow \citep{Lozano2021} and flow over 
 a realistic aircraft model in landing configuration deploying high-lift devices  \citep{goc2021large}.


Although WMLES is now gaining popularity as a high-fidelity tool balancing the computational cost and the accuracy, with the potential to be used for design and optimization in practical engineering applications because of its reasonable turnaround times, comprehensive benchmark studies on the comparison of different wall models are lacking. \citet{Park2017} compared the performance of ODE equilibrium wall model and PDE nonequilibrium wall model in a separating and reattaching flow over the NASA wall-mounted hump. \citet{LozanoDuran2020} tested three RANS based wall models (ODE equilibrium wall model, intergral nonequilibrium wall model and PDE nonequilibrium wall model) in a three-dimensional transient channel flow. For the latter study, it is worth noting that the three wall models were not tested using the same LES code. The lack of a like-for-like comparison of different wall models, especially in flows with nonequilibrium effects such as mean flow three-dimensionality and pressure gradient, warrants a systematic study of various wall models under the identical settings of the same solver and the same flow conditions. This will facilitate a clear assessment of the differences in the performance of different wall models, both in terms of the accuracy and the computational cost. Foregoing in view, in the present work, we test three wall models in a three-dimensional turbulent boundary layer (3DTBL) flow: an ODE equilibrium wall model (EQWM), an integral nonequilibrium wall model (integral NEQWM), and a PDE nonequilibrium wall model (PDE NEQWM). These three models respectively represent increasing model complexity with correspondingly increasing physical fidelity for predicting 3DTBLs. The equilibrium wall model assumes that the velocity profile is unidirectional and neglects all nonequilibrium effects, while the latter two are capable of representing skewed velocity profiles and incorporate some or all nonequilibrium effects, albeit in an averaged sense.

 Before describing the 3DTBL in more detail, a few remarks are in order regarding the suitability of the current choice of 3DTBL flow to conduct the comparative study of wall models with different physical fidelity. Historically, much of the research on wall turbulence has focused on statistically two-dimensional (2D) equilibrium turbulence in simple geometries (e.g., channel, pipe and flat plate). Different wall models perform equally well therein, making it hard to justify the use of more complex models. Furthermore, many practical flows of interest, such as those found on the swept wings of aircraft, wing/body juncture, bow/stern regions of ships and turbomachinery, are strongly affected by the mean flow three-dimensionality. Such 3DTBLs challenge the validity of the theories and models established from the canonical 2D wall turbulence and thus provide a good stage to exhibit the distinctive capabilities of different wall models. Therefore, the current study of turbulent boundary layer with mean-flow three-dimensionality is well suited to test different wall models, and to explain the physical origins of the differences in the results of these models.
 
The 3DTBLs can be classified as pressure-driven (also termed skew-induced \citep{Bradshaw1987} or inviscid-induced  \citep{LozanoDuran2020}),  or shear-driven (also termed viscous-induced \citep{LozanoDuran2020}) ones, according to the mechanisms by which the mean three dimensionality is introduced into the flow.
\color{black}
For the pressure-driven 3DTBLs, the crossflow is induced by the imposition of spanwise pressure gradient. 
More specifically, the mean three-dimensioanlity is produced by reorienting (tilting) the existing mean spanwise vorticity to generate non-zero streamwise vorticity. This process is often referred to as ``inviscid skewing'' due to its quasi-inviscid nature, and streamwise variation in the imposed spanwise pressure gradient often facilitates this vorticity tilting \citep{Coleman2000}. 
\color{black}
Examples of this type of 3DTBLs include flows in a square duct with a bend \citep{Schwarz1994,Flack1994}, in an S-shaped duct \citep{Bruns1999}, over wing-body junctures \citep{Rumsey2018}, over swept wings \citep{Bradshaw1985}, and over prolate spheroids \citep{Chesnakas1994}. 
For the shear-driven 3DTBLs, the crossflow is induced by the viscous diffusion of mean spanwise shear from the wall. Examples of this class include flows within a spinning cylinder \citep{Bissonnette1974,Lohmann1976,Driver1988}, over a rotating disk \citep{Littell1993}, over turbomachinery and in Ekman layers. 
\color{black}
In the present work, we are interested in the \textit{skew-induced} cases which are more prevalent in external hydrodynamics or aerodynamics applications.   

Over the past decades, studies on 3DTBL have unraveled its distinctive features which set it apart from the canonical 2D wall turbulence. First, the mean-flow direction in 3DTBL varies along the wall-normal direction, resulting in a skewed velocity profile. The law-of-the-wall, which is the characteristic of the canonical 2D wall turbulence, is therefore challenged in 3DTBL. Second, the Reynolds shear stress vector is not aligned with the mean velocity gradient vector in 3DTBLs. Thus the Reynolds shear stress response in 3DTBL can lag behind or lead that predicted by the isotropic eddy viscosity models which assume perfect alignment of the two. Third, a reduction in the structure parameter (the ratio of the total Reynolds shear stress magnitude to twice the turbulent kinetic energy) is often observed in 3DTBLs, whereas this parameter is nearly constant (roughly 0.15) in the outer layer of 2DTBL. The aforementioned features of 3DTBL pose a fundamental challenge to the validity of the underlying assumptions in many turbulence models (including wall models) that are based on 2DTBL, and therefore bring into question the reliability of these models when applied to practical flows. 

The numerical studies of 3DTBLs using direct numerical simulation (DNS) and large-eddy simulation (LES) have mostly focused on deformed 2D wall turbulence. These studies include channel flow subject to sudden crossflow pressure gradients \citep{LozanoDuran2020,Sendstad1992}, channel flow with spanwise wall motions, channel flow subject to mean strains \citep{Coleman2000}, TBL over an idealized infinite swept wing generated by a transpiration profile \citep{Coleman2019}, TBL subject to streamwise-varying pressure gradient \citep{Bentaleb2013}, and TBL on a flat plate with a time-dependent free-stream velocity vector \citep{Spalart1989}. These numerical studies are limited to relatively low Reynolds number and idealized 3DTBLs due to the large computational cost. 
The present study focuses on a realistic, spatially-developing, pressure-driven 3DTBL over the floor of a duct with a bend \citep{Schwarz1994}, which is at a considerably higher Reynolds number than the past studies but still providing a good balance between the physical realism, the tractability of the underlying 3DTBL mechanisms, and the computational cost of the simulations. 

The major objective of this work is to evaluate the three aforementioned wall models in the 3DTBL. Another objective is to understand the characteristics of the skew-induced 3DTBL, especially compared with the viscous-induced 3DTBL. The paper is organised as follows. The computational details including the flow configuration, boundary conditions and wall-model formulations are discussed in section \ref{sec:numerics}. The flow statistics obtained from WMLES are presented in section \ref{sec:results}. Based on these results, the performance of different wall models are compared and the characteristics of this pressure-driven 3DTBL are discussed based on the anisotropy invariant map and the Johnston triangular plot. The effects of different nonequilibrium terms in wall models in terms of predicting near-wall flow direction are also quantified. Finally, conclusions are given in section \ref{sec:conclusion}.

%
\section{\label{sec:numerics}Computational Details}
\subsection{\label{sec:flow_config}Flow configuration}

The reference configuration for the present study is the experimental setup of \citet{Schwarz1994}. While numerous experimental studies have been reported on 3DTBLs (as discussed in \citet{Flackreview1996}), our choice of the reference experiment was motivated primarily by the following aspects of \citet{Schwarz1994} which we found to be favorable to the goals of this study: 1) the highest Reynolds number among the pressure-driven 3DTBLs experiments reported in \citet{Flackreview1996}, 2) availability of the mean velocity and full Reynolds stress profiles, and of 3) the skin friction (magnitude and orientation) and pressure distribution along the wall. 
However, some remarks are also in order regarding 
limitations of the experiment. 
First, direct wall-stress data is not available, instead, a fit to the log law near $y^+ \approx 100$ was used for indirect stress measurement. Second, the description of the zero-pressure gradient region
far upstream of the bend region 
 for the purpose of CFD inflow generation is incomplete, therefore requiring an iterative procedure in the inflow generation to match the reported statistics at the first streamwise measurement station. 
\color{black}

\begin{figure}
\centerline{\includegraphics[width=0.8\textwidth]{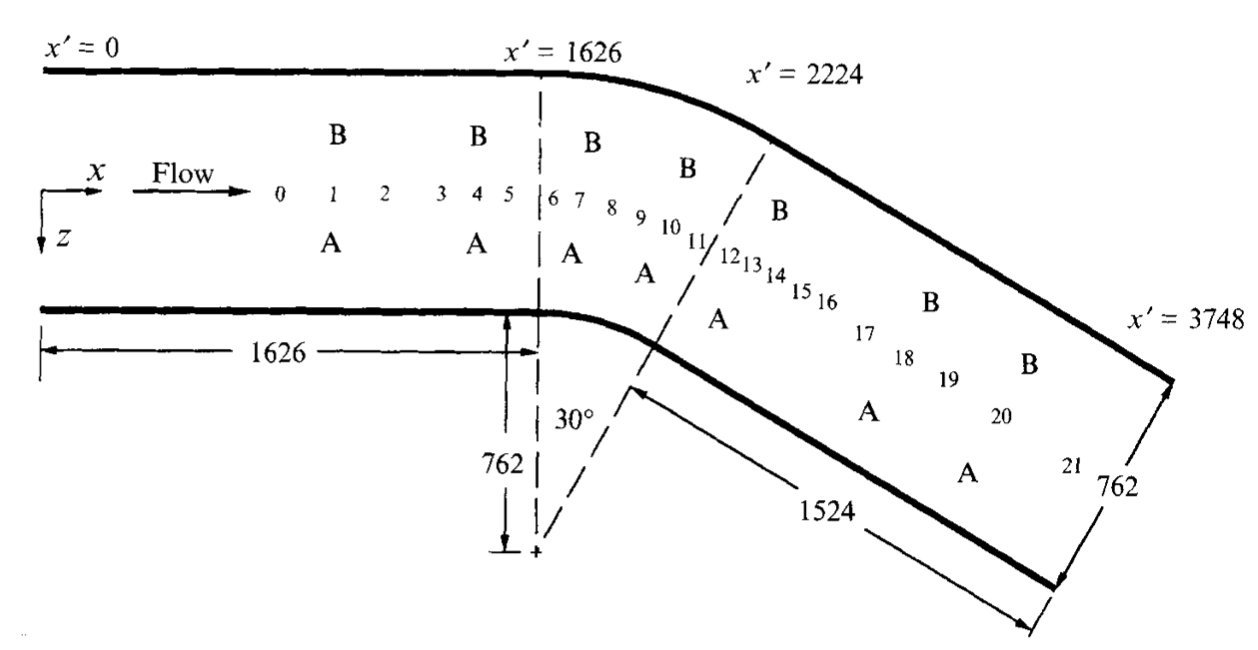}}
\caption{A Schematic of the floor of the duct (reproduction from figure 1 of \cite{Schwarz1994}). The measurement locations in the experiment are marked as numbers 0-21 along the duct centerline. Two coordinate systems are employed. $(x,y,z)$ is a fixed coordinate system with the origin located at the inlet. $(x',y',z')$ is a curvilinear coordinate system aligned with the local duct centerline (measurements in mm).}
\label{fig:duct}
\end{figure}

The experiment featured a spatially developing incompressible turbulent boundary layer, growing along the floor of a square duct with a 30$^{\circ}$ bend (see Fig.~\ref{fig:duct}). It should be noted that the boundary layer on the floor was very thin compared to the duct height, with $\delta_{99}/D$ ranging between 0.026 and 0.07 throughout the test section, where $D$ is the width (or height) of the square duct. The flow was far from being fully developed, and the secondary flow near the corner regions was expected to have negligible influence on the centerline region, which was the primary region of interest in the experiment. 

Following \citet{Schwarz1994}, two coordinate systems are employed here to facilitate the presentation of the results: $(x,y,z)$ denotes the global Cartesian coordinate system; $(x',y',z')$ denotes a curvilinear coordinate system aligned with the local duct centerline. $y = y'$ are the wall-normal coordinates (distance from the floor of the duct). In the experiment, the boundary layer on the floor was tripped using a trip wire at the duct inlet located at $x'=0$, thus ensuring a turbulent boundary layer over the entire floor of the test section. Boundary layers on the other three walls of the duct were not tripped (Schwarz, private communication, 2019). Reynolds number is moderately high, with $Re_{\theta}$ ranging between 4100 and 8500 (or $Re_{\tau}$ roughly ranging between 1500 and 3900). The flow along the centerline upstream of the bend was reported to exhibit typical characteristics of the canonical 2D zero pressure gradient (ZPG) flat-plate boundary layer. Mean flow three-dimensionality was generated in the bend region approximately between $x'=1626$ mm and $x'=2224$ mm due to the cross-stream pressure gradient induced by the bend. The surface streamlines were deflected by up to 22 degrees relative to the local duct centerline. Downstream of the bend, the 3DTBL gradually returned to a 2DTBL owing to the vanished spanwise pressure gradient. The experimental study focused on the boundary layer along the local centerline where the streamwise pressure gradient was found to be small.

\begin{figure}
\centering
\includegraphics[width=1.0\textwidth]{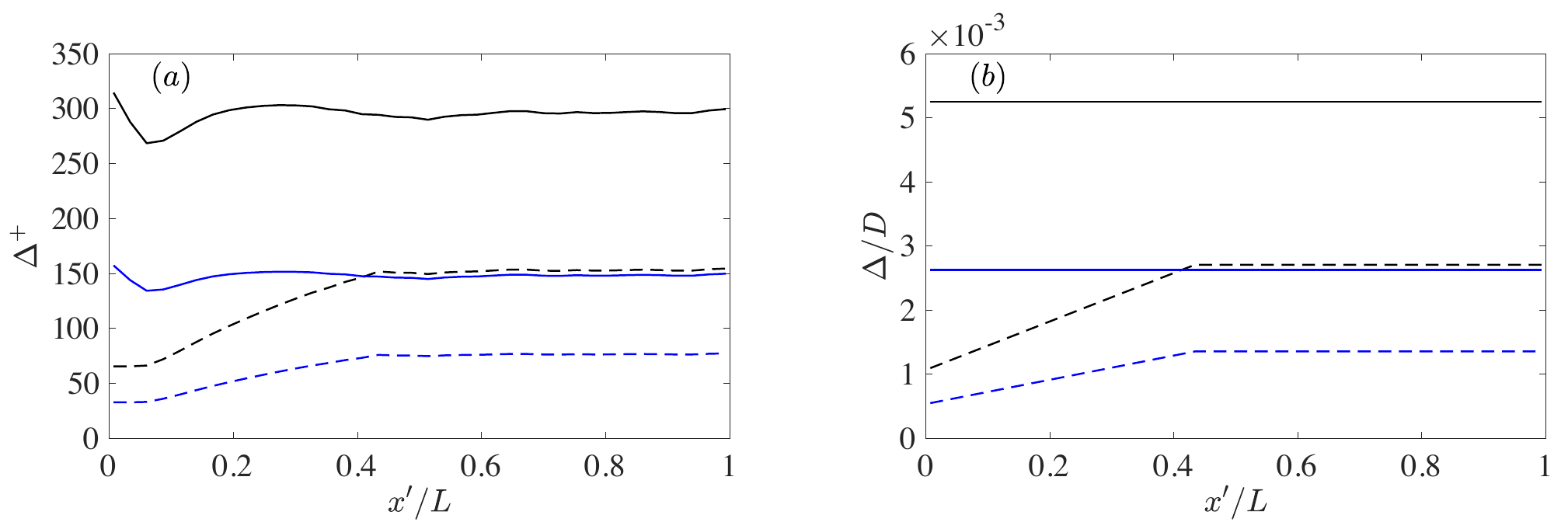}
\caption{\label{fig:grid_spacing} ($a$) Near-wall grid spacing distributions in wall units (based on the local skin friction) along the duct centerline. ($b$) Near-wall grid spacing distributions (normalized by duct height $D$) along the duct centerline. Solid lines, wall-parallel grid spacing ($\Delta x$ = $\Delta z$); dashed lines, wall-normal grid spacing ($\Delta y$). Black, coarse mesh; blue, fine mesh.}
\end{figure}

The computational domain is identical to the test section in the experiment, which consisted of a square duct ($D\times D=$ 0.762m $\times$ 0.762m) with a total curved length of $L$ = 3.748m, as shown in Fig.~\ref{fig:duct}. Two grid resolutions are considered in the present study: a coarse mesh with 8 million control volumes and a fine mesh with 38 million control volumes. Figure ~\ref{fig:grid_spacing} shows the near-wall grid spacing distributions along the duct centerline in the two meshes. 
\color{black}
The computational meshes are designed to maintain adequate wall-modeled LES grid resolution in the test section such that the local boundary layer contains approximately 16$\sim$23 and 32$\sim$45 cells across its thickness in the coarse and fine computational meshes, respectively. Local grid adaptations were applied in the near-wall region with the effect that the grid resolution transitions from the coarser isotropic-cell region in the freestream ($\Delta  = 0.008$ m at $y/D >$ 0.1) toward the finer near-wall region on the duct floor through anisotropic grid refinements. 
This resulted in wall-parallel grid resolutions ($\Delta x$ = $\Delta z$) of 4 mm and 2 mm for the coarse and fine meshes, respectively. 
In the region upstream of the bend ($x'/L \le 0.43$)  where the boundary-layer was thin but grew fast, 
the wall-normal grid spacing were varied with $x'$ to keep the number of boundary-layer resolving cells approximately constant, resulting in $\Delta y$ = 0.86 mm$ \sim$ 2.2 mm and $\Delta {y}$ = 0.43 mm $\sim$ 1.1 mm for the coarse and fine meshes, respectively. At $x'/L \ge 0.43$, $\Delta y$ was  fixed at their maximum values aforementioned.  
Compared to \citet{Cho2021} where WMLES of the same geometry using isotropic voronoi cells was reported, the present study using anisotropic hexahedral  cells deploys roughly the same wall-normal resolutions and about twice coarser wall-parallel resolutions in and downstream of the bend. Total cell counts are significantly reduced as a result, while maintaining higher numbers of cells across the thickness of the local boundary layer. 
\color{black}
The grid-resolution transition zones are located sufficiently away from the shear layer on the duct floor,  so that the solution therein is not affected by the accuracy degrade associated with abrupt changes in grid resolution. 

\subsection{\label{sec:citepref}Inflow characterization and boundary conditions}
\begin{figure}
\centerline{\includegraphics[width=1.0\textwidth]{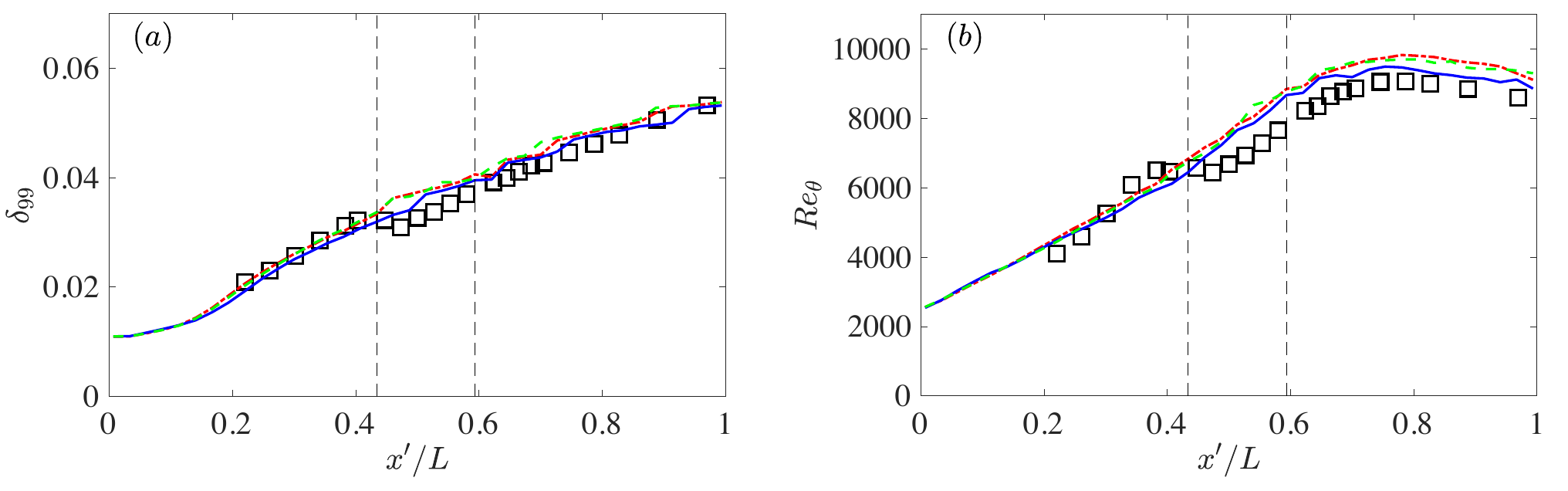}}

\caption{Centerline distributions of ($a$) boundary layer thickness and ($b$) momentum thickness (coarse mesh). Symbols, experiment; red dash-dotted line, equilibrium wall model; blue solid line, PDE nonequilibrium wall model; green dashed line, integral nonequilibrium wall model. Black vertical dashed lines denote the start and end of the bend region.}
\label{fig:integral_para}
\end{figure}

\begin{figure}
\centerline{\includegraphics[width=0.7\textwidth,trim={0 0 0 2cm},clip]{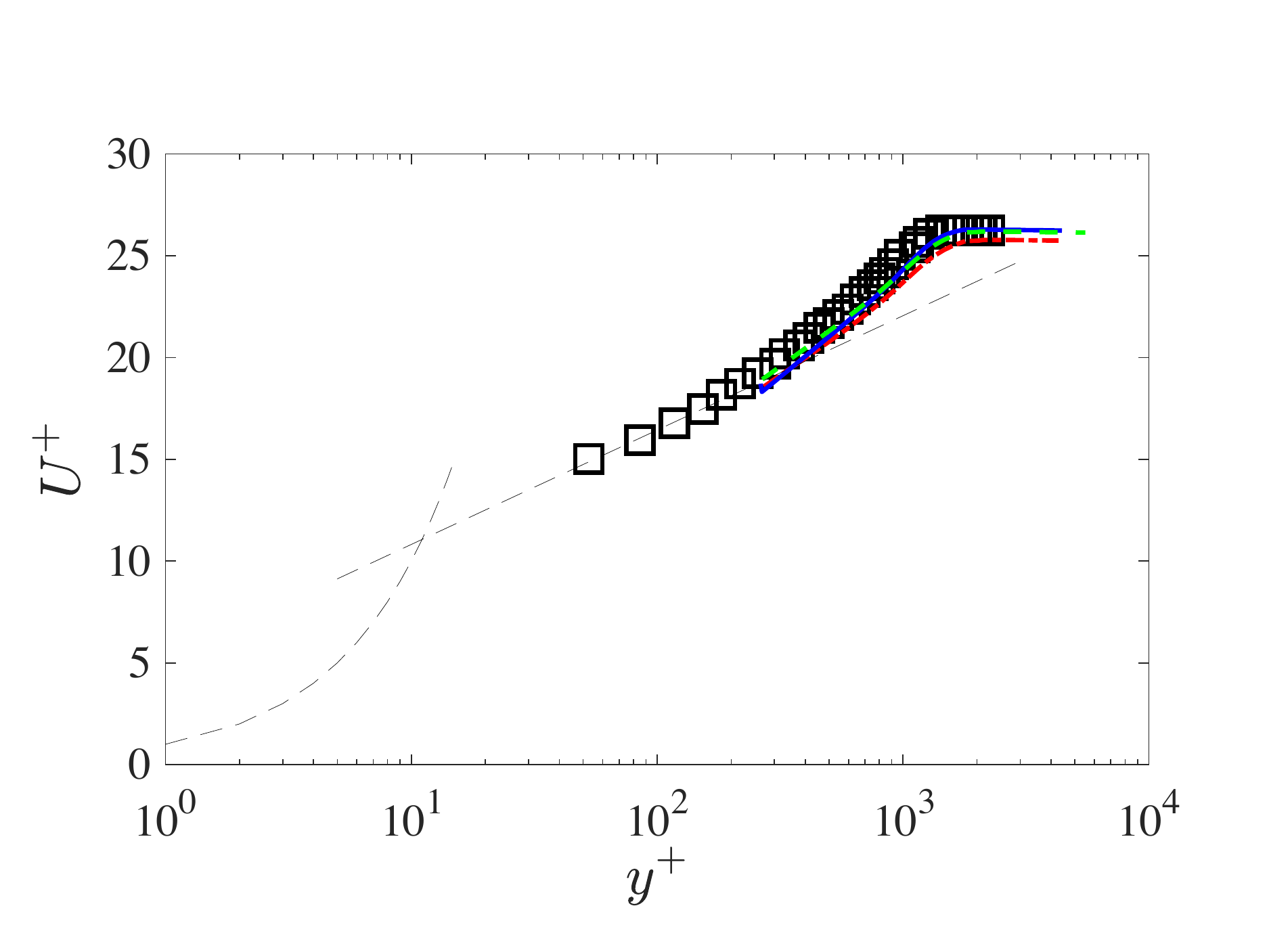}}
\caption{\label{fig:velocity} Mean velocity profile at the first measurement station in the experiment (station 0: $x'= 826$ mm) (coarse mesh). Symbols, experiment; red dash-dotted line, equilibrium wall model; blue solid line, PDE nonequilibrium wall model; green dashed line, integral nonequilibrium wall model. Black dashed lines denote the law of the wall (viscous sublayer: $u^+ = y^+$; log-law: $u^+ = \frac{1}{0.41}\log y^++5.2$).}
\end{figure}

Setting the appropriate boundary conditions in the simulation, particularly for the reproduction of flow characteristics upstream of the bend region where a typical equilibrium 2DTBL is expected, is crucial before attempting to compare the simulation results with the experimental results at any downstream location. However, the experiment reports flow statistics at the 22 locations shown in Fig.~\ref{fig:duct} along the duct centerline, with the first measurement location being far downstream of the test section inlet (at $x'=826$ mm). In the absence of this critical flow information at $x'=0$ mm, where the boundary layer on the floor was tripped in the experiment, we resort to a synthetic turbulence generation based on a digital filter approach \citep{Klein2003} for approximating the inflow boundary condition, rather than trying to replicate the trip-wire transition in the  experiment. This approach requires iterative guesses on the length of the development region (if any) to be appended upstream of the nominal trip location in the experiment ($x'=0$ mm), and the state of the inflow to be prescribed at the new inlet location. It should be noted that the goal here is to reproduce the 2DTBL upstream of the bend reasonably well, which then acts as the inflow for the 3DTBL within the bend, rather than to exactly match the flow conditions at the test section inlet. After iterating on several inflow conditions and the inlet location, 
we found that prescribing a flat-plate turbulent boundary layer at $Re_{\theta}=2560$ \citep{Schlatter2010} at the nominal inlet ($x'=0$ mm) reproduces the boundary layer statistics well at the first measurement location (station 0: $x'=826$ mm). As shown in Fig.~\ref{fig:integral_para}, the simulation agrees well with the experiment in terms of the distributions of the boundary layer and momentum thicknesses. 
In Fig.~\ref{fig:velocity}, the mean velocity profile at the first measurement station (station 0: $x'=826$ mm) is also shown to be reproduced well in the present calculations.

The prescription of boundary conditions on the rest of the boundaries is relatively straightforward. A subsonic Navier-Stokes characteristic boundary condition \citep{Poinsot1992} is imposed at the outlet of the duct. No attempt was made to resolve the boundary layers on the two side walls and the top wall which were not tripped in the experiment. 
The no-slip boundary condition is applied to each of these walls. The wall model is applied to the bottom wall, and the wall stress calculated from the wall-model solution is used as the Neumann boundary condition on this wall. All walls are assumed to be thermally adiabatic. 

\subsection{\label{sec:citeref}Flow solver and SGS / near-wall modeling}
The simulations were performed with CharLES, an unstructured cell-centered finite-volume compressible LES solver developed at Cascade Technologies, Inc. The solver employs an explicit third-order Runge-Kutta (RK3) scheme for time advancement and a second-order central scheme for spatial discretization. More details regarding the flow solver can be found in \cite{Khalighi2011} and \cite{Park2016}. The Vreman model \citep{Vreman2004} is used to close the SGS stress and heat flux. 

\begin{figure}
\centerline{\includegraphics[width=1.0\textwidth,trim={0 0 0 0cm},clip]{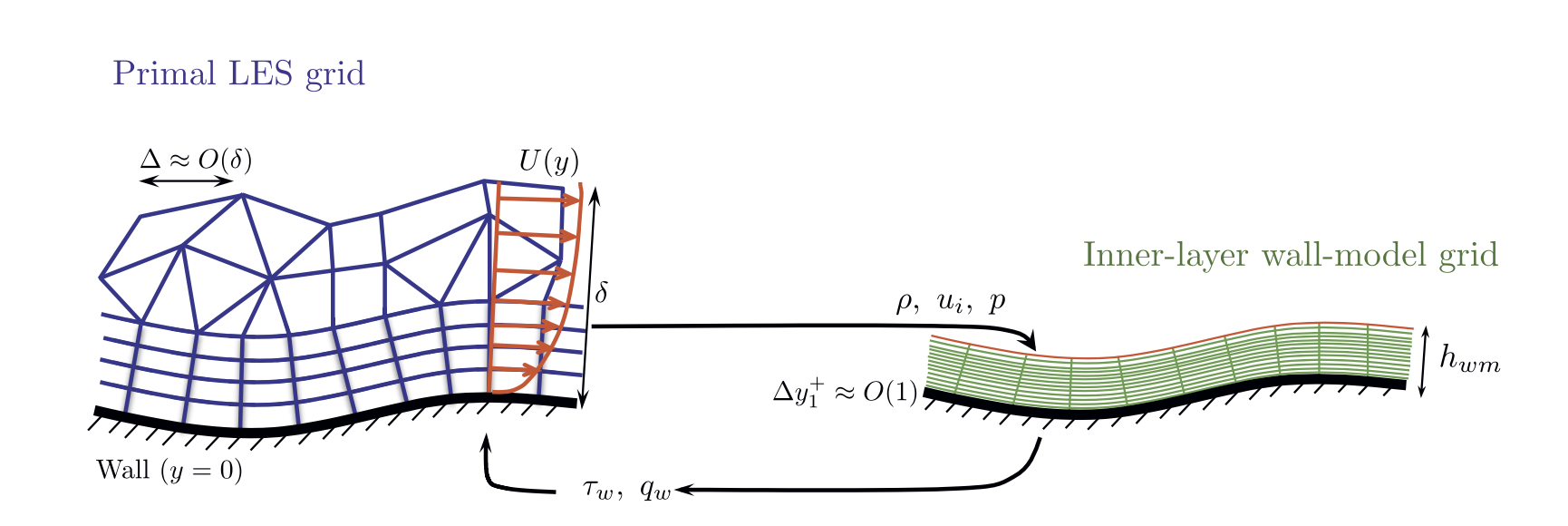}}
\caption{\label{fig:wm_schematic} Sketch of the wall-modeling procedure (reproduction from figure 1 of \cite{Park2016}). Wall shear stress ($\tau_w$) and heat flux ($q_w$) are solved from the wall model equations on a separate near-wall mesh. Wall models are driven by the LES states imposed at their top boundaries while the no-slip condition is applied at the wall.  }
\end{figure}

In WMLES, LES equations are solved on a coarse mesh, where the stress-carrying eddies in the near-wall region are mostly unresolved. The LES mesh alone cannot represent the sharp velocity gradients and the momentum transport near the wall. This causes SGS models to produce insufficient levels of modeled stresses. Wall modeling aims to compensate for such numerical and modeling errors in the underresolved near-wall region of LES, by augmenting the total stresses directly through the imposition of the modeled stress boundary condition at the wall in lieu of the no-slip condition. In the present work, wall models solve simplified, vertically integrated, or full RANS equations on a separate near wall mesh. The grid for wall models have fine resolution in the wall-normal direction (with the exception of the integral nonequilibrium wall model, which does not require a wall-normal grid), but the wall-parallel grid resolution (if any) is identical to or coarser than the LES grid. All wall models in this study are driven by the LES states imposed at their top boundaries, which are taken at a specified matching height in the LES grid. At each time step, wall stress and heat flux obtained from the wall model are used as the Neumann wall boundary condition for the LES. Figure~\ref{fig:wm_schematic} shows a schematic of the wall modeled LES procedure employed in the present work.

In the present work, the 
location at which the wall-models take input from the LES 
(matching height, denoted as $h_{wm}(x)$) is fixed across different grid resolutions by setting it to the centroids of the third off-wall cells or the top faces of the fifth off-wall cells in the coarse and fine meshes, respectively, corresponding to $10^3 h_{wm} / D  =  6  \ \sim  \ 14$. This has effect of fixing the wall-modeled regions in LES during grid refinements, so that improvement in LES prediction is 
not associated with 
 the change in  wall-modeling details, but it is attributed largely to the grid adaptation.  This choice is also motivated by our experience with 
 the flow solver, 
 where 
restricting $h_{wm}$ to the first off-wall cell or in the buffer layer 
produced nontrivial log-layer mismatch in channel flow calculations, even with the filtration of the wall-model input as suggested in \citet{Yang2017} for a structured pseudospectral/finite-difference code. \citet{Owen2020} reported a similar need for using the LES velocity further away from the wall in their finite-element based WMLES of channel and wall-mounted hump. 
\color{black}

The three wall models considered in the present study are: an equilibrium stress model (EQWM) in the form of ordinary differential equations (ODE), an integral nonequilibrium wall model (integral NEQWM) that solves the vertically-integrated  Navier-Stokes equations, and a PDE nonequilibrium wall model (PDE NEQWM) that retains the complexity of the full Navier-Stokes equations. All three wall models parameterize the unresolved turbulence in the wall-model domain in a statistical sense using simple RANS models based on the mixing-length formulation. Note that the EQWM and PDE NEQWM have previously been implemented in CharLES, and they were tested extensively through various studies \citep{Park2014,Park2016,Park2016PRF,Park2017,Bodart2012conference}. 
The  integral NEQWM was  recently integrated into CharLES, the implementation aspects of which will be discussed in a future article \citep{Hayat2021}. A brief description of each of these models is given below.

The EQWM \citep{Kawai2012,Bodart2011} solves the simplified boundary layer equations which account only for the wall-normal diffusion.
\begin{eqnarray}
    \frac{d}{d\eta}\left[(\mu + \mu_{t})\frac{du_{||}}{d\eta}\right] = 0,\\
    \frac{d}{d\eta}\left[(\mu + \mu_{t})u_{||}\frac{du_{||}}{d\eta}+(\lambda + \lambda_{t})\frac{dT}{d\eta}\right] = 0,
\end{eqnarray}
where $\eta$ is the local wall-normal coordinate, $u_{||}$ is the wall-parallel velocity magnitude, $T$ is the temperature, $\mu$ is the molecular viscosity, $\lambda$ is the molecular thermal conductivity, and $\mu_t$ and $\lambda_t$ are the turbulent eddy viscosity and conductivity, respectively. The velocity vector is assumed to be aligned with the LES velocity at the matching height. Owing to this intrinsic assumption, the equilibrium wall model is incapable of predicting skewed velocity profiles within the wall-modeled domain. 
Also, due to the unidirectionality and the condition $\mu+\mu_t > 0$, the EQWM can represent monotonic velocity profiles only, and it 
cannot predict velocity profiles with  sign changes in the slope as found in the near-wall regions of separated shear layers. 
\color{black}
The wall-model eddy viscosity $\mu_t$ is based on the following mixing-length formula,
\begin{eqnarray}
    \mu_{t} = \kappa \rho y\sqrt{\frac{\tau_w}{\rho}}D,\  D=[1-{\rm exp}(-y^+/A^+)]^2.
\end{eqnarray}

On the other hand, the PDE NEQWM \citep{Park2014,Park2016} solves the full 3D unsteady RANS equations,
\begin{eqnarray}
\frac{\partial \rho}{\partial t} + \frac{\partial \rho u_{j}}{\partial x_{j}}=0,\\
\frac{\partial \rho u_{i}}{\partial t} + \frac{\partial \rho u_{i}u_{j}}{\partial x_{j}} + \frac{\partial p}{\partial x_{i}}=\frac{\partial \tau_{ij}}{\partial x_{j}},\\
\frac{\partial \rho E}{\partial t} + \frac{\partial(\rho E+p)u_{j}}{\partial x_{j}}=\frac{\partial \tau_{ij}u_{i}}{\partial x_{j}}-\frac{\partial q_{j}}{\partial x_{j}},
\end{eqnarray}
where $\rho$ is the density and $u_{i}$ is the velocity component, $p$ is pressure and $E=p/[\rho(\gamma-1)]+u_{k}u_{k}/2$ is the total energy. The stress tensor and heat flux are given by, $\tau_{ij}=2(\mu+\mu_t)S_{ij}^d$ and $q_{j}=-(\lambda + \lambda_{t})\frac{\partial T}{\partial x_{j}}$. For the RANS closure, a novel mixing-length model is used, which dynamically accounts for the resolved Reynolds stresses carried by the wall model \citep{Park2014}.
The wall-model mesh for the PDE NEQWM has the same wall-parallel grid content as in the coarse LES mesh, but it is refined in the wall-normal direction to resolve the viscous sublayer. 

\color{black}
Lastly, the integral NEQWM  formulation solves a similar set of equations as the PDE NEQWM, albeit in a wall-normal integrated form. Currently, this formulation is limited to incompressible flows, and therefore the energy equation is not solved. For the sake of brevity, only the 2D formulation (the wall-normal and one wall-parallel velocity components) is presented below,  and the reader is referred to \citet{Yang2015} for the details of full 3D formulation. The vertically integrated momentum equation is given by,
	\begin{equation}\label{eq:IWM}
	\frac{\partial}{\partial t}\int_{0}^{h_{wm}} u dy + 
	\frac{\partial }{\partial x} \int_{0}^{h_{wm}} u^2 dy -
	U_{LES} \,\frac{\partial}{\partial x} \int_{0}^{h_{wm}} u d y = 
	\frac{1}{\rho} \left[-\frac{\partial p}{\partial x} h_{wm} 
	+\tau_{h_{wm}} - \tau_{w}\right],
	\end{equation}
where $x$ and $y$ represent the local wall-parallel and wall-normal coordinates, $h_{wm}$ is the matching height, $U_{LES}$ is the time-filtered velocity from the LES solution at the matching location.   
$\tau_{w} = \mu \left.\frac{\partial u}{\partial y}\right|_{y=0}$ and 
$\tau_{h_{wm}} = \left(\mu+\mu_{t}\right)\frac{\partial u}{\partial y}\big |_{y=h_{wm}}$ are the shear stresses at the wall and at the matching location, respectively. 
The integral terms are evaluated by assuming an analytical composite profile for the velocity within the wall model, which has the form:

    \begin{eqnarray}
    u = u_{\tau} \frac{y}{\delta_{\nu}} =  \frac{u_{\tau}^2}{\nu}y, & \,\,\,\, 0 \leq y \leq \delta_{i}, \\
    u = u_{\tau}\left[\frac{1}{\kappa} \log \frac{y}{h_{wm}}+C\right]+u_{\tau} A \frac{y}{h_{wm}}, & \,\,\,\, \delta_{i}<y \leq h_{wm},\label{eq:linear_departure}
    \end{eqnarray}
where the unknown parameters $A$, $C$, $u_{\tau}$ and $\delta_{i}$ are determined from the solution of equation (\ref{eq:IWM}) along with suitable matching and boundary conditions. For the full 3D formulation consisting of two wall-parallel velocity components, like the one employed for the present study, the composite profiles have a total of 11 unknown parameters. 
This approach attempts to model the effects of pressure gradient and advection through the last term in Eq.~(\ref{eq:linear_departure}) representing linear departure from the log law.  
\color{black}

It is worth mentioning here that in the original 3D formulation in \citet{Yang2015}, the assumed form of the viscous-sublayer velocity profiles in the two wall-parallel directions (Eq. (C5) in \citet{Yang2015}) resulted in inconsistent asymptotic behavior of velocity near the wall. This made the wall-stress predictions of the wall model highly sensitive to the choice of the local $x$/$z$ coordinates. In our current integral NEQWM formulation, we modify the assumed viscous-sublayer profile to ensure consistent near-wall asymptotic behavior as given by the Taylor series expansion. The details of this modified formulation along with its implementation in an unstructured solver are presented in \citet{Hayat2021}. The MATLAB implementation for the EQWM and the integral NEQWM are available on GitHub at \url{https://github.com/imranhayat29/Wall-Models-for-LES}.

A remark is in order regarding the overall cost of simulations with different wall models. The computational costs of the three wall models were compared by running the simulations on the fine LES mesh with 256 CPU cores for three convective flow-through times.  When the cost of the simulation without any wall model (no-slip LES) is taken as the unity, the simulation costs are 1.27, 1.2, and 2.2 with the EQWM, the integral NEQWM, and the PDE NEQWM, respectively. The higher cost with the PDE NEQWM is due to the inversion of a large linear system required as a part of implicit time advancement. 
\color{black}

\section{\label{sec:results}RESULTS}
Results from the WMLES simulations are discussed in this section. Overall characteristics of the flow are first highlighted from the instantaneous flow field standpoint. Mean and turbulence statistics obtained with different wall models are then assessed against the experimental data. Furthermore, the three-dimensionality of the outer portion of the boundary layer is examined with the aid of Reynolds stress anisotropy and the Johnston triangular plot. 

Some remarks are in order concerning the ways the main results are presented in this paper. The primary interest of the experiment was to examine the effect of the mean three dimensionality in the absence of strong streamwise pressure gradient. To this end, the experiment presented the key flow statistics along the floor centerline, where the axial pressure gradient was observed to be nearly zero. It should be noted, however, that the mean-flow trajectory near the wall deviates somewhat substantially from the centerline as the flow passes through the bend region, as observed from the instantaneous flow field (Fig.~\ref{fig:isosurface_Q}) and the near-wall streamlines (Fig.~\ref{fig:flow_direction}($b$)). This leaves some ambiguity in the interpretation of the statistics presented along the centerline,  because any two fluid particles on the centerline (in and after the bend region) would have traveled along different Lagrangian trajectories, experiencing different history effects (most notably, they are subject to different upstream axial/spanwise pressure gradients.) With this limitation in mind, we still choose to show our results along the centerline, as all experimental data (except the wall-pressure) were presented so. 

\subsection{\label{sec:level2}Grid convergence}
\begin{figure*}
\centerline{\includegraphics[width=1.0\textwidth]{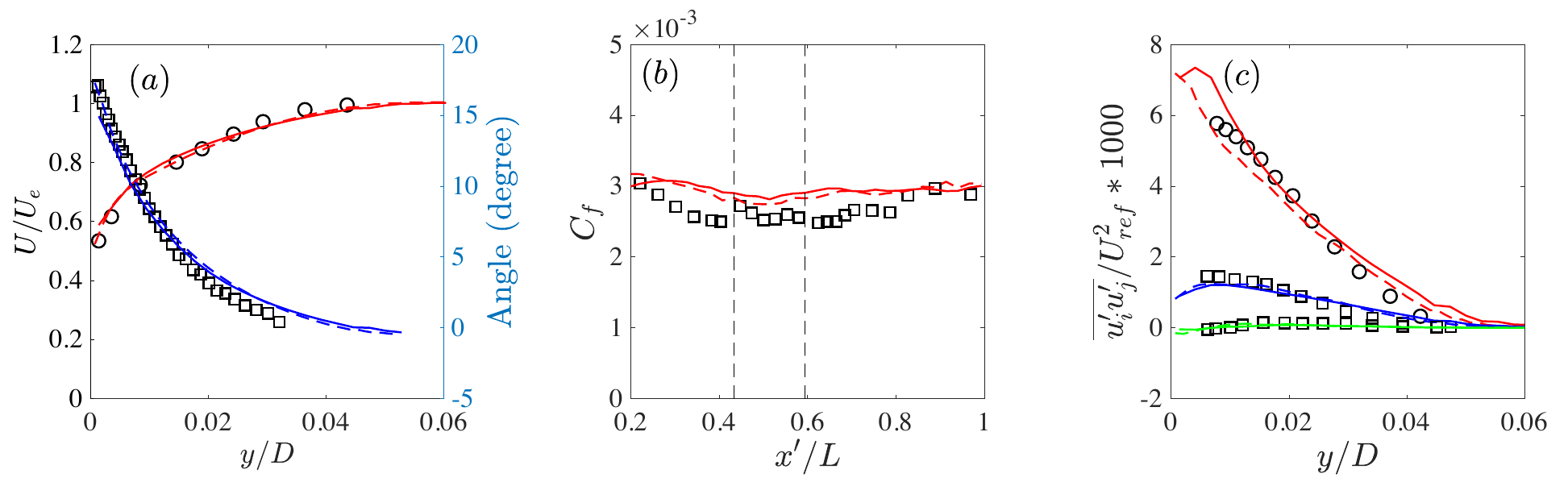}}
\caption{\label{fig:grid_convergence} Grid convergence of the EQWM LES. ($a$) Blue, mean flow direction vs. distance from the wall; red, mean velocity magnitude profile, ($b$) Centerline distribution of the skin-friction coefficient, and ($c$) Reynolds-stress profiles. Red, $\overline{u'u'}$; blue, $-\overline{u'v'}$; green, $\overline{v'w'}$. Solid lines, coarse grid; dashed line, fine grid. Profiles in ($a$) and ($c$) are at $x'=1875$ mm (station 8). Black vertical dashed lines in ($b$) denote the start and end of the bend region.}
\end{figure*}

Figure~\ref{fig:grid_convergence} shows the mean flow statistics and the Reynolds stresses at $x'=1875$ mm (the eighth measurement station in Fig.~\ref{fig:duct}), as well as the centerline $C_{f}$ distribution obtained from the EQWM LES with the coarse and the fine grids described in Sec.~\ref{sec:flow_config}. Although only the EQWM results are shown here for brevity, it is noted that the other two wall models exhibited similar grid-convergence characteristics. In  Fig.~\ref{fig:grid_convergence}($a$), both the mean velocity and the mean flow direction (see Sec.\ref{sec:mean_flow_stats} for definition) profiles obtained from the coarse-grid calculation are already in good agreement with the experiment, and the results only improve marginally with the grid refinement. More importantly, this points toward the grid convergence of the results for the refinement level used in this study. Figure ~\ref{fig:grid_convergence}($b$) shows the skin-friction distribution along the centerline of the duct. Between the first and the last measurement stations, we observe a reasonably converged $C_{f}$, with the fine-grid calculation  producing slight improvement in $C_{f}$. In Fig.~\ref{fig:grid_convergence}($c$), a similar trend  is observed for all the Reynolds stress components shown, with the exception of the streamwise component of the Reynolds normal stress, which shows noticeable variation with grid refinement in the near-wall region; however, the Reynolds stresses in the outer portion of the boundary layer have largely converged. Having established reasonable evidence of grid-convergence for most of the flow statistics on the coarse grid, the remainder of this paper will focus largely on discussing the results obtained with the coarse grid unless stated otherwise.
\color{black}
\subsection{\label{sec:level2}Instantaneous flow field}
\begin{figure*}
\centerline{\includegraphics[width=1.0\textwidth]{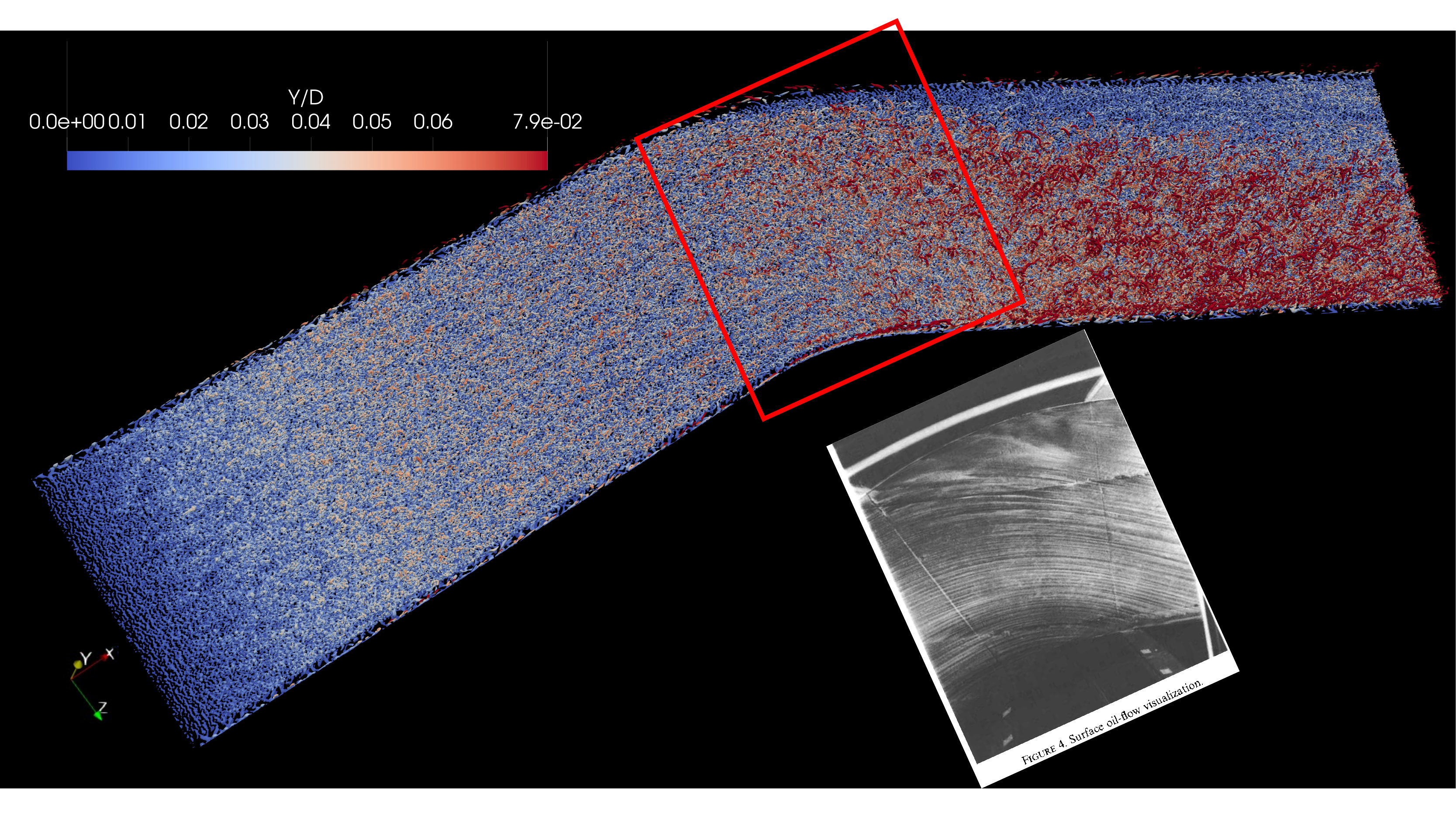}}
\caption{\label{fig:isosurface_Q} Visualization of the near-wall vortical structures using isosurface of $Q$ (the second invariant of the velocity gradient tensor), colored by the distance from the floor of the duct. Flow is from the bottom left to the top right. Surface oil visualization from the experiment \citep{Schwarz1994} is shown in the inset (figure reprinted with permission from \cite{Schwarz1994}.}
\end{figure*}

Figure~\ref{fig:isosurface_Q} visualizes the vortical structures in the floor boundary layer. Near the inlet (approximately within 20 times the inlet boundary-layer thickness from the inlet), structures with less coherence resulting from the synthetic inflow turbulence generation are observed. The floor boundary layer then gradually develops into a coherent fully developed state far upstream of the bend region, which is also verified by the velocity profile at $x'=826$mm (Fig.~\ref{fig:velocity}) following the typical law of the wall observed in 2D turbulent boundary layer. When the flow enters the bend region, a clear contrast of two boundary layers with different origins (blue and red regions) are observed. The boundary layer in the red region is thicker than that in the blue region, showing that a new boundary layer is emerging from the concave sidewall (at $z<0$) and the original boundary layer developed from the upstream section is turning toward the convex side. This overall flow behavior is visually in fair agreement with the surface oil visualization in the experiment as shown in the inset in Fig.~\ref{fig:isosurface_Q}. 

\subsection{\label{sec:mean_flow_stats}Mean flow statistics}
\begin{figure*}
\centerline{\includegraphics[width=0.8\textwidth,trim={0 3cm 0 3cm},clip]{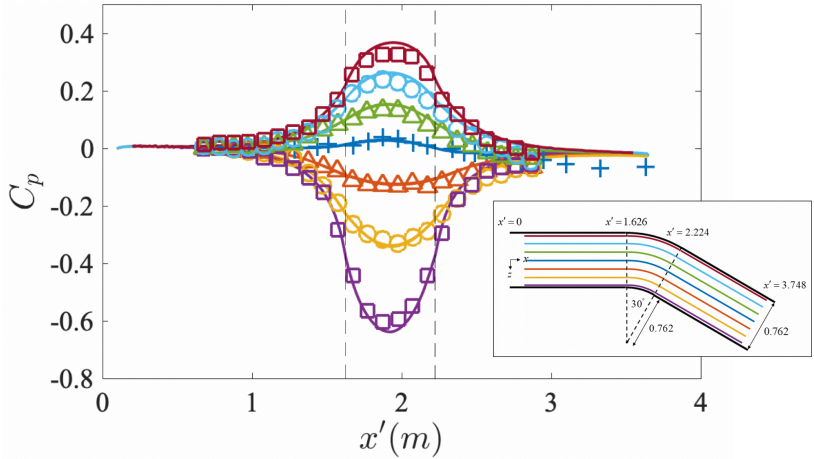}}
\caption{\label{fig:Cp} Variation of the wall-pressure coefficient from coarse EQWM simulation (results from the other wall models and resolutions are almost identical). Colors denote different spanwise locations corresponding to the inset figure. $z':0$ mm, $\pm 127$ mm, $\pm 254$ mm, $\pm 368$ mm }
\end{figure*}

The cross-stream pressure gradient is the source of the mean three dimensionality in the bend region. It acts to deflect the streamlines close to the wall more strongly than those near the free stream. 
It is therefore important to first establish close agreement 
in the pressure distribution close to the bend 
between the simulation and the experiment.  
Note that in the current case without flow separation, the pressure distribution is determined largely by the wall geometry and the inviscid effect, presumably unaffected by the wall-modeling details.  
Figure~\ref{fig:Cp} shows the distribution of the static wall-pressure coefficient on the floor of the duct. The pressure coefficient is defined as
\begin{equation}
    C_{p} = \frac{p-p_{ref}}{\frac{1}{2}\rho U^2_{ref}}.
\end{equation}
Following the experiment, $p_{ref}$ is the static pressure at $x'=0$ mm, and $U_{ref}$ is the freestream velocity at $x'=826$ mm (defined at the spanwise centerline). 
The wall-pressure probing lines are parallel to the duct centerline as shown in the inset figure. The figure shows good agreement between the simulation and the experiment, except in the recovery region downstream of the bend. 
The axial pressure gradient is almost zero along the centerline. 
On the other hand, a significant spanwise pressure gradient starts to develop upstream of the bend, reaches a maximum within the bend region, and eventually decays to zero downstream of the bend. 
The reason for disagreement in the recovery region remains unclear to us.
While $C_p$ in the experiment remains to be slightly negative near the outlet, 
$C_p$ in the simulation naturally vanishes to its upstream zero value as the flow relaxes back to its 2D ZPG state. 
Note that the experiment reported only the centerline distribution in this region,  and that extending the duct further  downstream in the simulation did not change the trend.

\begin{figure}
\centerline{\includegraphics[width=1.0\linewidth]{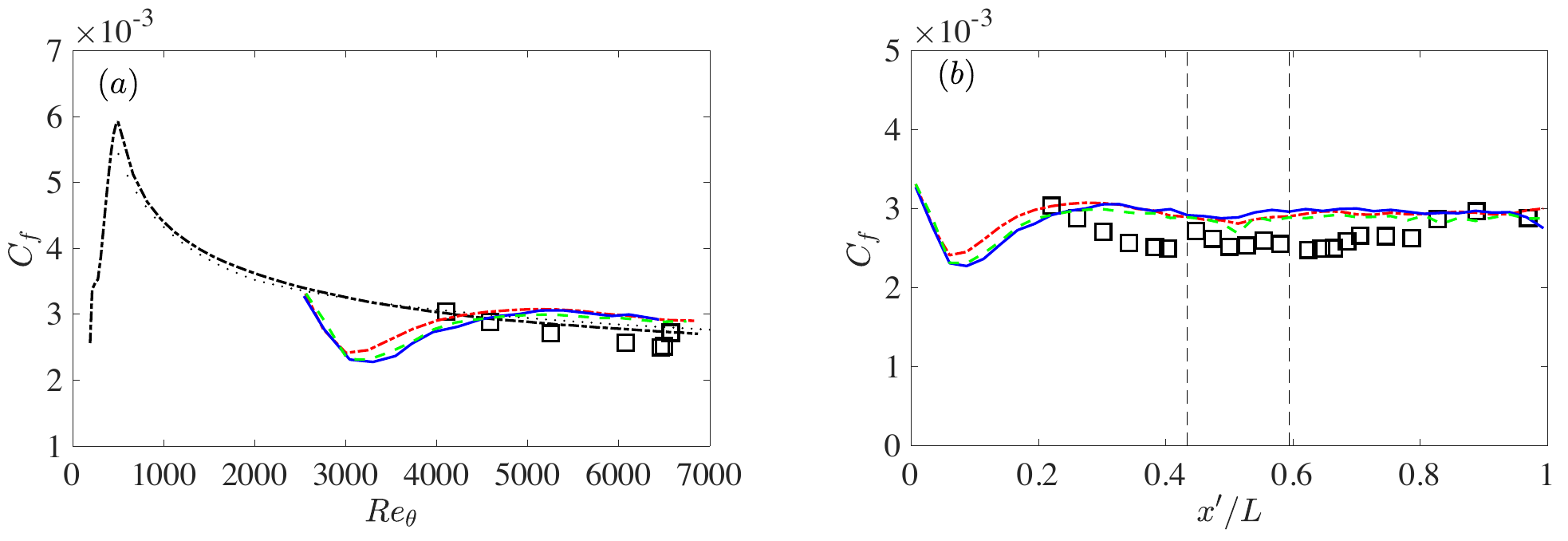}}

\caption{Centerline distribution of the skin-friction coefficient ($C_f$). ($a$) $C_f$ vs. $Re_{\theta} $ upstream of the bend. ($b$) $C_f$ vs. axial location. Squares, experiment; red dash-dotted line, EQWM; blue solid line, PDE NEQWM; green dashed line, integral NEQWM. 
In ($a$): 
Black dotted line,  zero pressure gradient flat-plate boundary layer (ZPGFPBL) empirical correlation (Eq.~(9) in \cite{Fernholz1996}); 
black dashed line, wall-resolved LES of ZPGFPBL  \citep{Schlatter2014}. 
} 
\label{fig:skin_friction}
\end{figure}

Figure~\ref{fig:skin_friction} shows the distribution of the skin-friction coefficient along the duct centerline. The skin-friction coefficient is  defined as, 
\begin{equation}
    C_{f}=\frac{\tau_{w}}{\frac{1}{2}\rho U^2_{e}} \ \ \ ,\quad \frac{U_e}{U_{ref}}=(1-C_{p})^{1/2},
\end{equation}
where $\tau_w = \sqrt{\tau_{w,x}^2+\tau_{w,z}^2}$ is the magnitude of the mean wall-shear stress vector, and $U_e$ is the local free-stream velocity. Figure~\ref{fig:skin_friction}($a$) shows the centerline distribution of the skin friction  upstream of the bend as a function of the momentum thickness Reynolds number. 
The centerline mean flow is expected to be agreeing well with 
the canonical ZPG 2DTBL in this region.  
A deviation of the skin friction from the ZPG 2DTBL near the inlet is the artifact of the inflow treatment. Note that the synthetic inflow 
turbulence generation 
methods when applied to DNS or wall-resolved LES of low Reynolds number 
are known to produce 
a development length of  $10 \sim 20$ initial boundary layer thicknesses ($\delta_{in}$), 
through which coherence-lacking artificial structures mature into fully-developed turbulence  (e.g., \cite{Patterson2021,Sandberg2012,Larsson2021}
where $Re_\tau = 400 \sim 500$). 
The present high-Reynolds number case simulated with very coarse meshes produced  longer development lengths (30$\sim$40 $\delta_{in}$). The flow was observed to be fully developed from slightly upstream of the first measurement station, after which the WMLES results are in reasonable agreement with the experiment as well as with an  empirical correlation \citep{Fernholz1996} and  a wall-resolved LES of ZPG 2DTBL \citep{Schlatter2014}.  In  Fig.~\ref{fig:skin_friction}($b$), slight overprediction of the skin friction from WMLES is observed throughout the duct. A similar trend was reported by \citet{Cho2021}, where the EQWM was used with up to 76 million control volumes in LES. It should be noted that the wall shear stresses were measured indirectly in the experiment using a Preston tube and Patel's calibration \citep{Patel1965} for a 2DTBL. \cite{Patel1965} reported 
that errors as large as 6\% could occur when Preston tube is used in flows with moderate streamwise pressure gradient. 
\color{black}

\begin{figure}
\centerline{\includegraphics[width=0.8\linewidth]{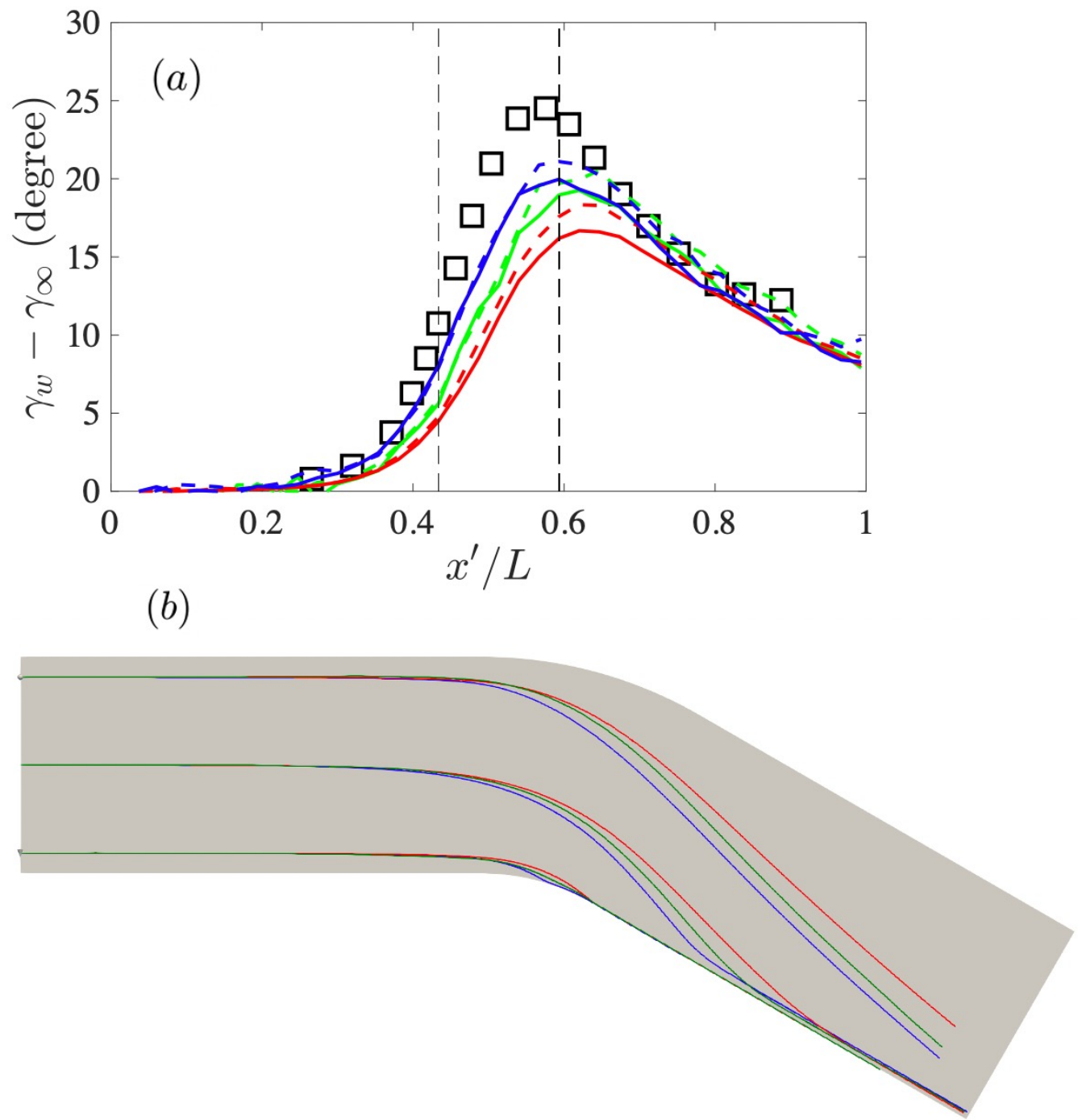}} 

\caption{($a$) Centerline distribution of the surface flow turning angles with respect to the freestream ($\gamma_w = \text{tan}^{-1}(\tau_{w,z}/\tau_{w,x})$ is the wall shear stress direction, $\gamma_{\infty} =  \text{tan}^{-1}(W_e/U_e)$ is the freestream direction). 
($b$) streamlines of wall shear stress. 
Squares, experiment; 
red line, equilibrium wall model; 
blue line, PDE nonequilibrium wall model; 
green line, integral nonequilibrium wall model. 
Solid and dashed lines are for coarse and find grids, respectively. 
}
\label{fig:flow_direction}
\end{figure}
Next, we examine the mean three-dimensionality of the flow in the duct. The variation of the surface flow direction relative to the freestream direction is a measure of the mean flow three-dimensionality. As  shown in Fig.~\ref{fig:flow_direction}($a$), the crossflow is almost zero in the upstream, and it grows rapidly as the flow approaches the bend. The resulting turning angle reaches the maximum near the end of the bend and decays gradually thereafter. These observations are consistent with the development of the spanwise pressure gradient. All three wall models predict the general trend in the turning-angle variation correctly; however, the PDE NEQWM gives the most accurate prediction among the three (especially within the bend region), followed by the integral NEQWM, and then the EQWM. The maximum difference between the PDE NEQWM and the EQWM is roughly 5 degrees occurring at $x'/L=0.49$ within the bend.
Note that the total flow turning is an accumulative effect of the local flow change, and the area under the curve in Fig.~\ref{fig:flow_direction}($a$) can be thought of as an approximation of the near-wall total flow turning angle. 
Related to this, fig.~\ref{fig:flow_direction}($b$) visually highlights predictions of the near-wall flow direction by different wall models in terms of the select surface streamlines calculated from the mean wall shear-stress vector. 
\color{black}
It can be clearly seen that the flow deviates from the local centerline; however, the deviation is not predicted evenly across the different wall models,  consistent with our observation in the flow turning angles in Fig.~\ref{fig:flow_direction}($a$). The surface flow from the PDE NEQWM turns much more rapidly than that from the EQWM. 
This has a direct implication for practical engineering flows involving 3DTBL. For instance, in the skew-induced 3DTBL over the swept wing of an aircraft, the circulation over the wing and the downwash characteristics could potentially be affected by the near-wall change in the flow direction, which may alter the lift. 
Additionally, errors in the surface-flow directionality can alter three-dimensional separation patterns observed near the wing-body juncture of wings at high-lift configurations \citep{Evans2020} to increase uncertainty in prediction of maximum lift and the associated drag. 
\color{black}

\begin{figure*}
\centerline{\includegraphics[width=1.0\textwidth]{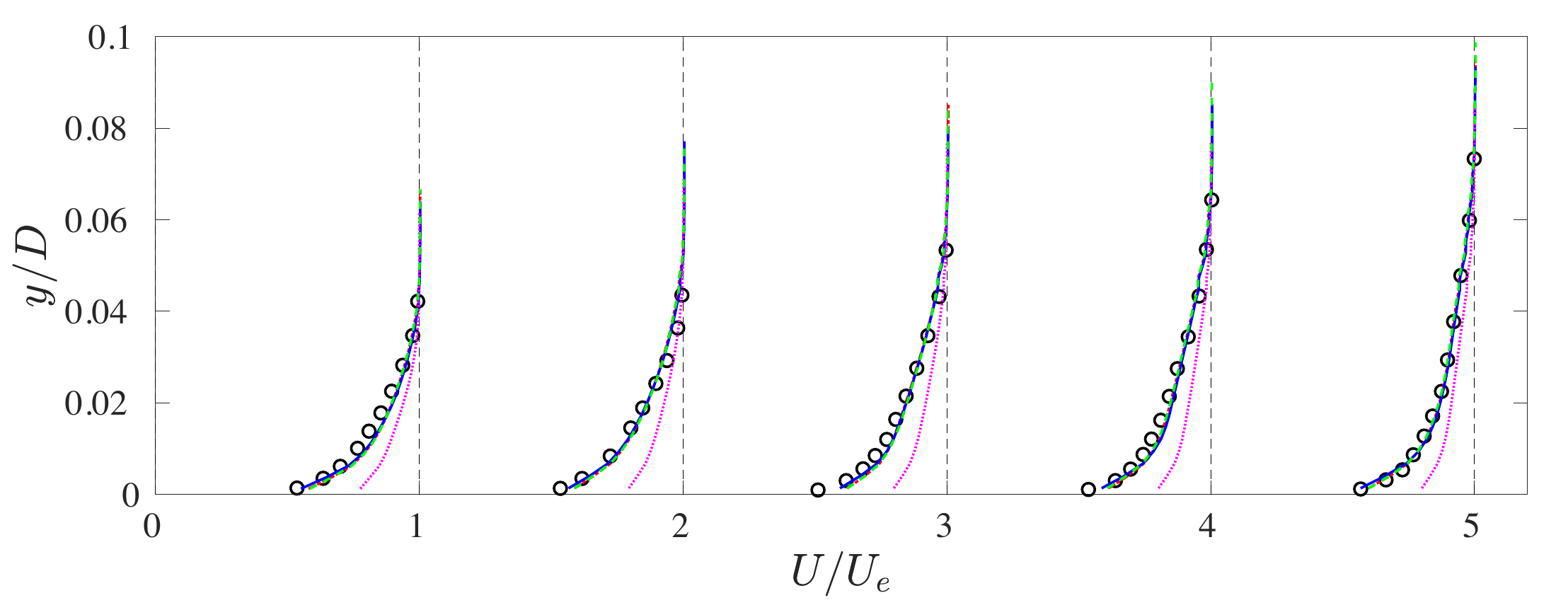}}
\caption{\label{fig:mean} 
Profiles of the mean-velocity magnitude at 5 measurement locations (stations 4, 8, 12, 16, and 20, from left to right).
Station 4 is upstream of the bend; station 8 is within the bend; stations 12, 16 and 20 are downstream of the bend. 
Red dash-dotted line, EQWM; 
blue solid line, PDE NEQWM; 
green dashed line, integral NEQWM; 
magenta dotted line, no-slip LES;
black circle, experiment. 
Profiles are shifted along the abscissa by 1. }
\end{figure*}
In Fig.~\ref{fig:mean}, profiles of the mean-velocity magnitude are compared between the different wall models and the experiment, at several locations along the centerline, including upstream of, within, and downstream of the bend. It can be seen that the no-slip LES, which does not employ a wall model, gives a very poor prediction of the mean velocity. Here, a higher momentum is imparted to the boundary layer as a consequence of the underpredicted wall shear force. 
\color{black}
With the introduction of wall modeling, a significant improvement is achieved in the predicted mean-velocity profiles. In line with the predictions of the skin-friction coefficient in Fig.~\ref{fig:skin_friction}, the mean velocity profiles across the three wall models are almost identical. Note that here, the profiles only show the magnitude of the mean velocity, thus lacking  information on the three-dimensionality of the mean flow.

\begin{figure}
\centerline{\includegraphics[width=0.7\textwidth,trim={0 0cm 0cm 0.5cm},clip]{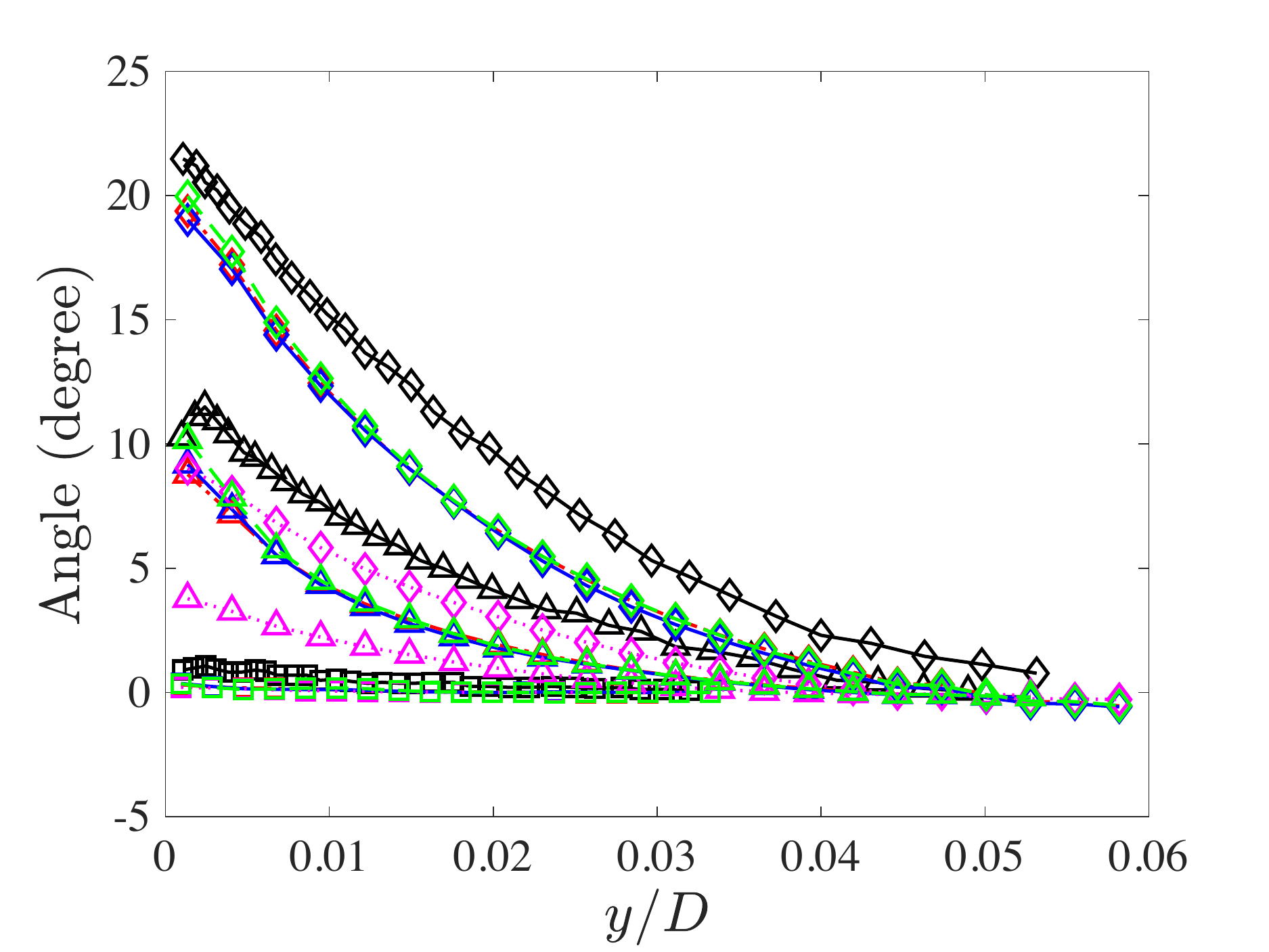}}
\caption{\label{fig:flow_angle_vs_y} Mean flow direction relative to the local freestream as a function of wall distance. 
Red dash-dotted line, EQWM; 
blue solid line, PDE NEQWM; 
green dashed line, integral NEQWM; 
magenta dotted line, no-slip LES; 
black solid line, experiment.
Symbols are used to differentiate stations along the duct floor centerline only (Square, station 0; triangle, station 6; diamond, station 10). Lines with the same symbols denote the results at the same stations.\color{black}}
\end{figure}

To complete this picture, Fig.~\ref{fig:flow_angle_vs_y} shows how the flow direction changes along the wall-normal direction. The flow direction characterizes the mean flow three-dimensionality and it is defined by the angle between the mean velocity vector and the freestream velocity vector. The mean flow three-dimensionality is the strongest at the wall and becomes weaker with the increase in distance from the wall, as evident from the diminishing crossflow away from the wall. A difference of approximately 3 degrees is observed between the WMLES and the experimental results. However, the predicted angles from the different wall models are almost identical, indicating that the difference in the wall-model outputs (the wall-shear force direction observed in Fig.~\ref{fig:flow_direction}($b$)) 
is not felt by the LES solutions away from the wall.

\subsection{\label{sec:level2}Reynolds stress}
We now turn our attention to the turbulent content of the 3DTBL and its role in distinguishing this flow from the canonical 2DTBL, by examining the Reynolds stress related statistics. Indeed, the Reynolds stresses in the 3DTBL exhibit unique characteristics not seen in the 2DTBL, as we will see shortly. 

\begin{figure*}
\centerline{\includegraphics[width=0.8\linewidth]{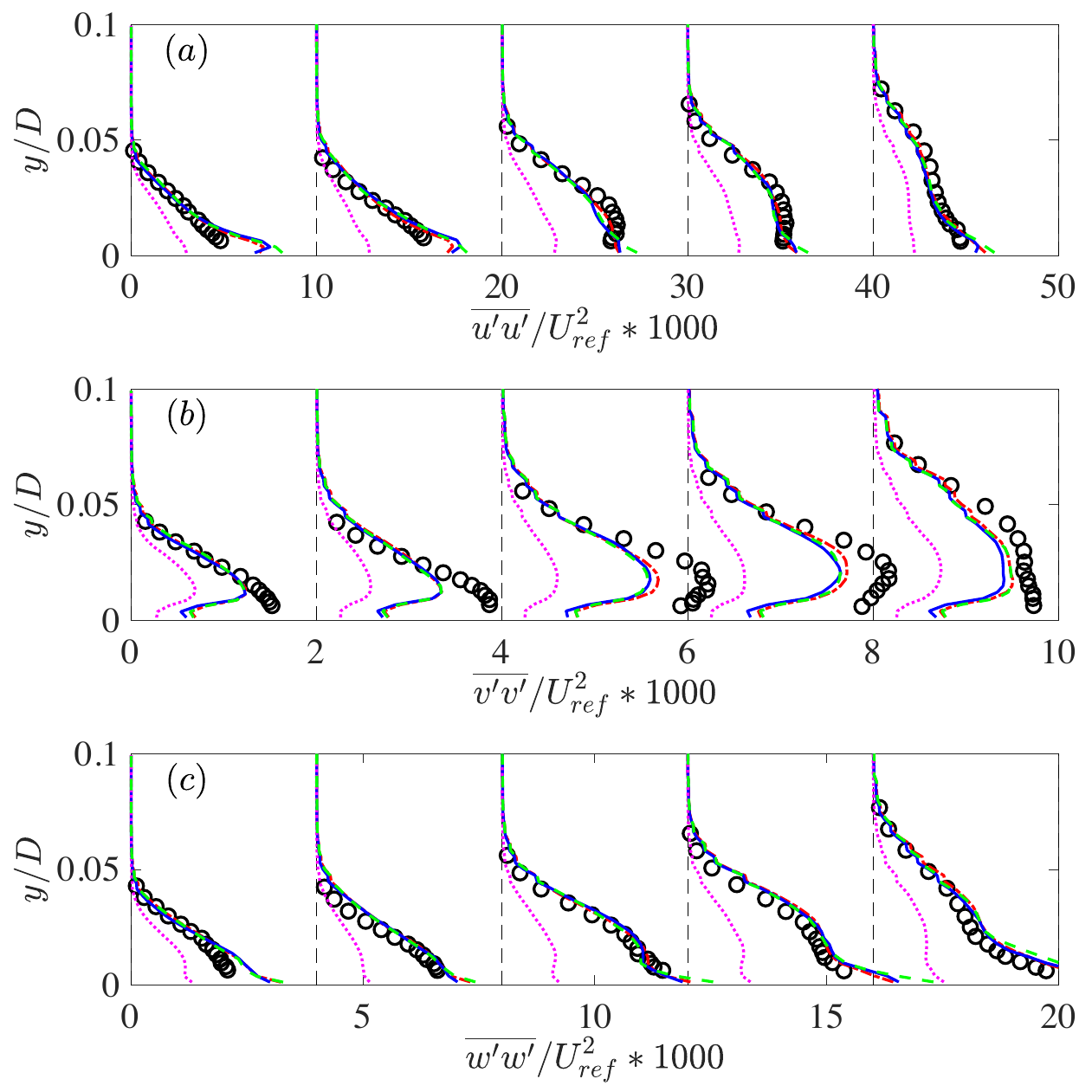}}

\caption{Reynolds normal stress profiles at the same five measurement stations as the mean velocity profiles in Fig.~\ref{fig:mean}. Red dash-dotted line, EQWM; blue solid line, PDE NEQWM; green dashed line, integral NEQWM; magenta dotted line, no-slip LES; black circle, experiment. Profiles are shifted along the abscissa by multiples of 10, 2 and 4 for $\overline{u'u'}$, $\overline{v'v'}$ and $\overline{w'w'}$, respectively.}
\label{fig:reynolds_normal}
\end{figure*}

\begin{figure*}
\centerline{\includegraphics[width=0.8\linewidth]{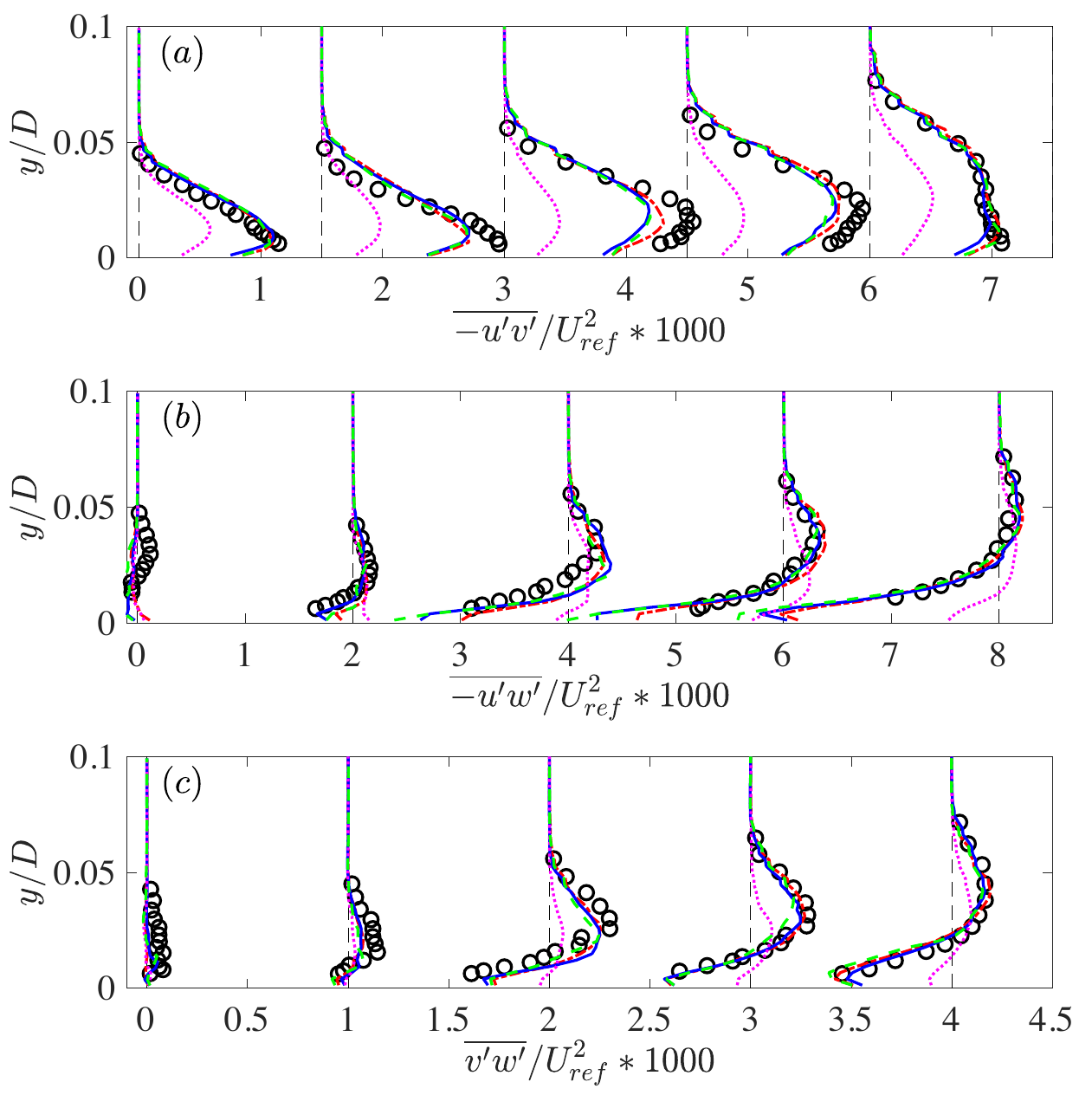}}

\caption{Reynolds shear stress profiles at the same five locations as the mean velocity profiles in Fig.~\ref{fig:mean}. Red dash-dotted line, EQWM; blue solid line, PDE NEQWM; green dashed line, integral NEQWM; magenta dotted line, no-slip LES; black circle, experiment. Profiles are shifted along the abscissa by multiples of 1.5, 2 and 1 for $\overline{-u'v'}$, $\overline{-u'w'}$ and $\overline{v'w'}$, respectively.}
\label{fig:reynolds_shear}
\end{figure*}

Figure~\ref{fig:reynolds_normal} shows the profiles of the Reynolds normal stresses at the same five centerline locations which were chosen to depict the mean velocity profiles. 
Note that the experiment did not have access to the Reynolds-stress data in the inner layer, therefore missing information on the peak values and their locations.
\color{black}
The no-slip LES acutely underpredicts the Reynolds normal stress, pointing towards the grid resolution being insufficient for the no-slip LES to resolve the near wall eddies. 
All three wall models significantly improve the prediction of the normal stresses, bringing the profiles closer to the experimental results. The predicted Reynolds normal stresses in the wall-parallel directions show remarkable agreement with the experiment, whereas those in the wall-normal direction are underpredicted near the wall. 
Figure~\ref{fig:reynolds_shear} shows the profiles of the Reynolds shear stresses, where substantial improvement with wall modeling is also observed. 

\begin{figure*}
\centerline{\includegraphics[width=1.00\textwidth]{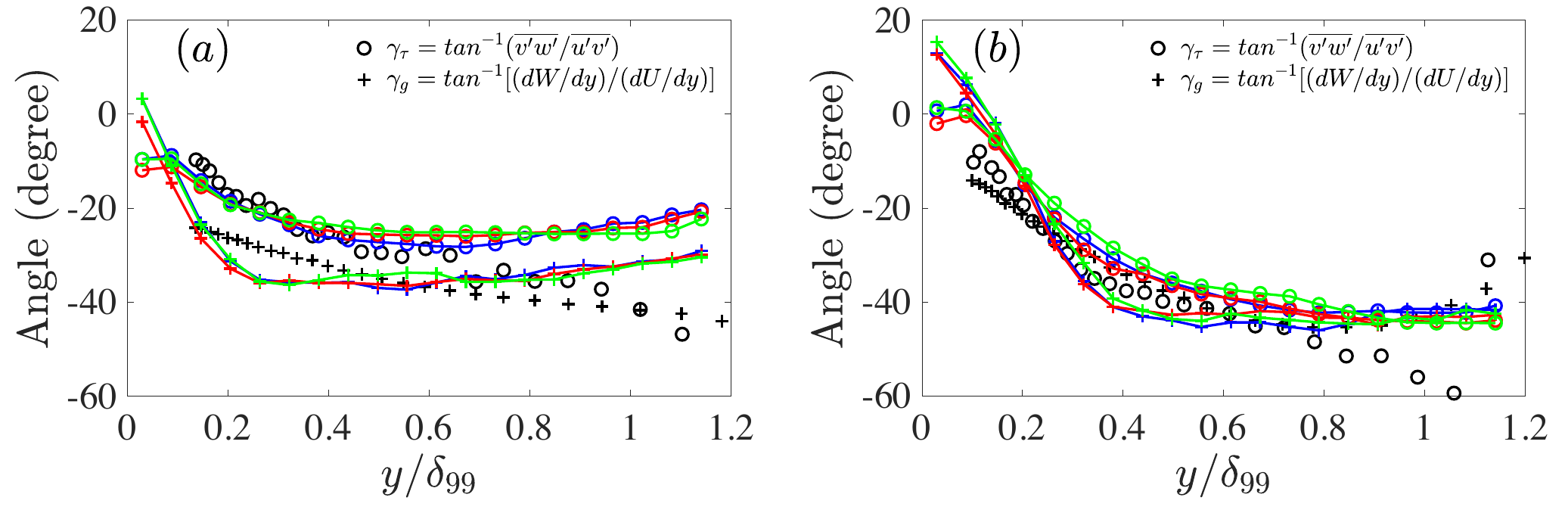}}
\caption{\label{fig:mean_shear_reynolds_shear} Directions (relative to the local freestream) of the mean velocity gradient vector and the Reynolds shear stress vector. ($a$) station 10; ($b$) station 17. Red, EQWM; blue, PDE NEQWM; green, integral NEQWM; black, experiment. Circle, angle between the Reynolds shear stress vector $\gamma_{\tau}$ and the local freestream; cross, angle between the mean velocity gradient vector $\gamma_{g}$ and the local freestream.}
\end{figure*}

An important characteristic of the 3DTBL is that the Reynolds shear stress vectors are not necessarily aligned with the mean velocity gradient vectors, which challenges the fundamental assumption of the commonly used isotropic eddy viscosity model. The directions of these two vectors are characterized by the angles defined below,
\begin{equation}
\gamma_{\tau}=\text{tan}^{-1}\Big(\frac{\overline{v'w'}}{\overline{u'v'}}\Big), \quad
\gamma_{g}=\text{tan}^{-1}\Big(\frac{\partial W/\partial y}{\partial U/\partial y}\Big). 
\end{equation}
Figure~\ref{fig:mean_shear_reynolds_shear}($a$) clearly shows that the Reynolds shear stress vector lags behind the mean velocity gradient vector within the bend where mean-flow three-dimensionality is strongest. Downstream of the bend where mean-flow three-dimensionality declines, the difference between the two vectors also decreases (Fig.~\ref{fig:mean_shear_reynolds_shear}($b$)). Furthermore, this lag appears to be a function of the distance from the wall. 
The experiment shows that the lag decreases with wall distance in the outer layer above $y/\delta_{99} \approx 0.7$ within the bend. Downstream of the bend, the Reynolds shear stress vector even starts to lead mean velocity gradient vector above $y/\delta_{99} \approx 0.8$, and this lead increases with wall distance. These behaviors and the shear-stress angles therein are not captured well in LES.  We conjecture that computation of the angles in this region is prone to contamination by numerical error or measurement noise, because  both the Reynolds shear stress and the mean velocity gradient values are very small there.  At $y/\delta_{99} \le 0.7$, a reasonable agreement between the simulations and the experiment is observed, and results from different wall models do not show notable difference.  
\color{black}

\subsection{\label{sec:level2}Lumley triangle: anisotropy invariant map}
We have noted that the Reynolds stresses from the simulations agree reasonably well with the experiment. This makes it possible to further investigate the anisotropy of the Reynolds stress in the outer layer of this 3DTBL  using the WMLES solution. In this section, using the fine grid prediction, we employ the Lumley triangle to analyze the Reynolds-stress anisotropy. Following \citet{Pope2000}, the normalized anisotropy tensor is defined as, 
\begin{equation}
    b_{ij} = \frac{\langle u_{i}u_{j} \rangle}{\langle u_{k}u_{k} \rangle} - \frac{1}{3}\delta_{ij}.
\end{equation}
The anisotropy tensor has zero trace and thus has  two independent principal invariants. It is convenient to define two variables $\xi$ and $\eta$, corresponding to the two invariants, as
\begin{eqnarray}
    6\eta^2=-2I_{2}=b^2_{ii}=b_{ij}b_{ji}\\
    6\xi^3=3I_{3}=b^3_{ii}=b_{ij}b_{jk}b_{ki}.
\end{eqnarray}
The state of anisotropy can be characterized by the above two variables $\xi$ and $\eta$. All realizable Reynolds-stress states must be located within a triangular region in the $\xi$-$\eta$ plane, which is known as the Lumley triangle.
The boundary of the Lumley triangle corresponds to some special states of the Reynolds-stress tensor: the origin corresponds to the isotropic turbulence; the left corner point corresponds to the two-component (2C) axisymmetric state; the right corner point corresponds to the one-component (1C) state; the left straight line corresponds to the axisymmetric contraction and the right straight line corresponds to the axisymmetric expansion; the top curve represents the two component (2C) turbulence.

\begin{figure}
\centerline{\includegraphics[width=1.0\linewidth]{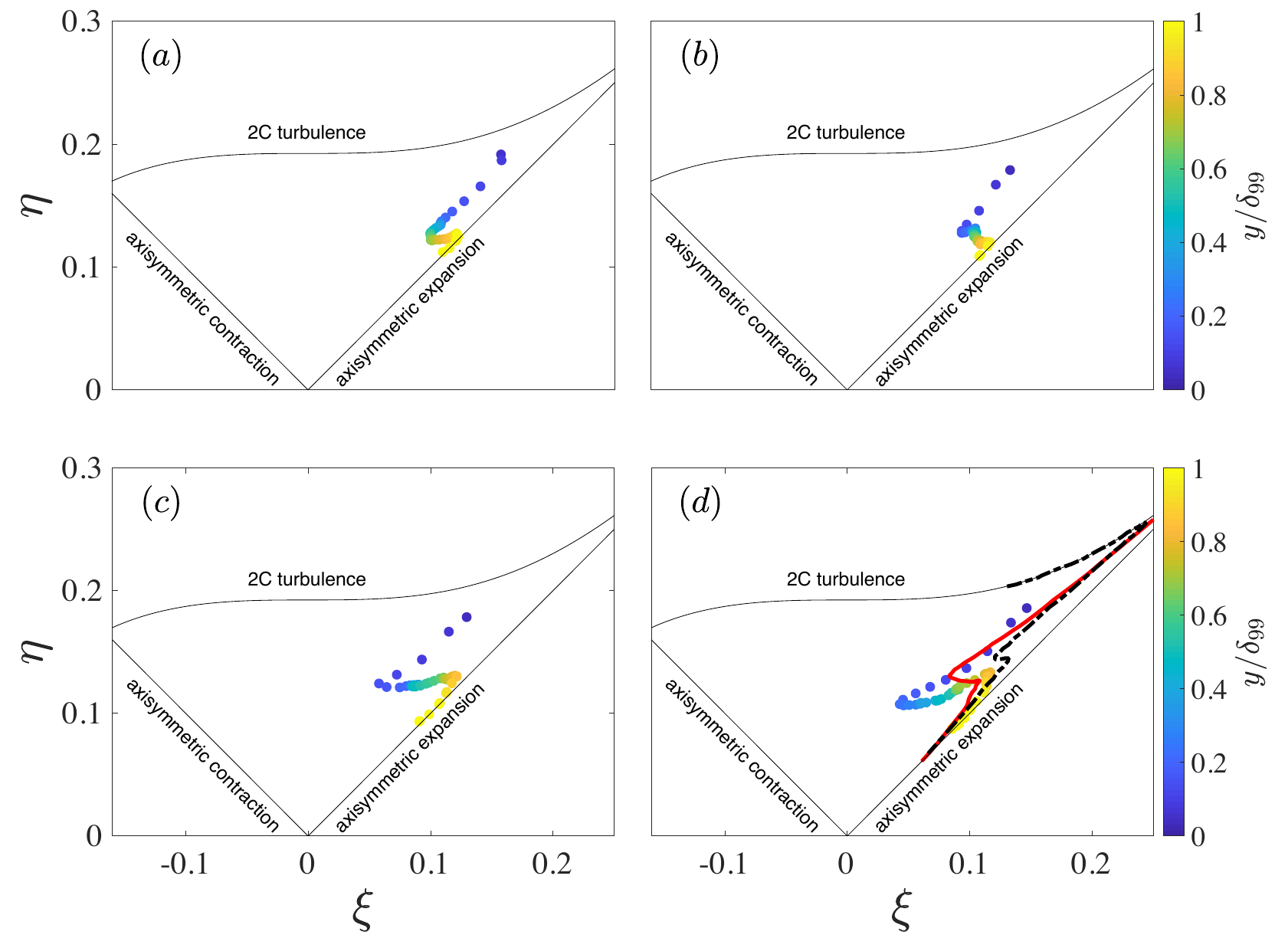}}

\caption{Lumley triangle of the WMLES results along the wall-normal direction.  ($a$) station 4. ($b$) station 10. ($c$) station 12. ($d$) station 18. 
Colored dots, PDE NEQWM result (fine grid); black dash-dotted line, canonical 2D channel flow at $Re_{\tau}=2003$ \citep{Hoyas2006}; red solid line, shear-driven 3DTBL from transient channel at $Re_{\tau}=546$ at $t^+=192$ \citep{LozanoDuran2020}. Colorbar denotes the wall distance normalized by the local boundary layer thickness.}
\label{fig:lumley_triangle}
\end{figure}

The evolution of the Reynolds stress anisotropies through the bend is shown in Fig.~\ref{fig:lumley_triangle}. 
Comparisons to a statistically 2D channel flow at $Re_\tau = 2003$ \citep{Hoyas2006} and a shear-driven transient statistically 3D channel flow at $Re_\tau = 546$ \citep{LozanoDuran2020} can be also made from these data shown in Fig.~\ref{fig:lumley_triangle}($d$). 
Right upstream of the bend, the wall-normal distribution of the anisotropies away from the wall shows some similiarty to that in the 2D channel (Fig.~\ref{fig:lumley_triangle}($a$)), exhiting a characteristic S-shape 
lying close to the axisymmetric-expansion (AE) limit. As the flow passes through the bend, the left cusp of this S curve rapidly dislocates toward inside the triangle, leaving less points close to the AE limit. The non-monotonic decrease of the anisotropies (with increasing wall distance) is also observed vividly, which is only weakly present in the 2D channel. While the station 18 is considerably downstream of the bend region, the anisotropies are still seen further departing from its 2D behavior. This is  consistent with the observation in Fig.~\ref{fig:mean_shear_reynolds_shear} that the Reynolds stresses respond 
more slowly than the mean to the imposed three-dimensionality. 
Also, in Fig.~\ref{fig:lumley_triangle}($d$), note the similarity of the anisotropy distributions in the duct and the shear-driven 3DTBL from the transient channel flow, although departure from the 2D behavior is much stronger in the duct case. 
\color{black}

\subsection{\label{sec:level2}Triangular plot}
\begin{figure}
\centerline{\includegraphics[width=1.0\linewidth]{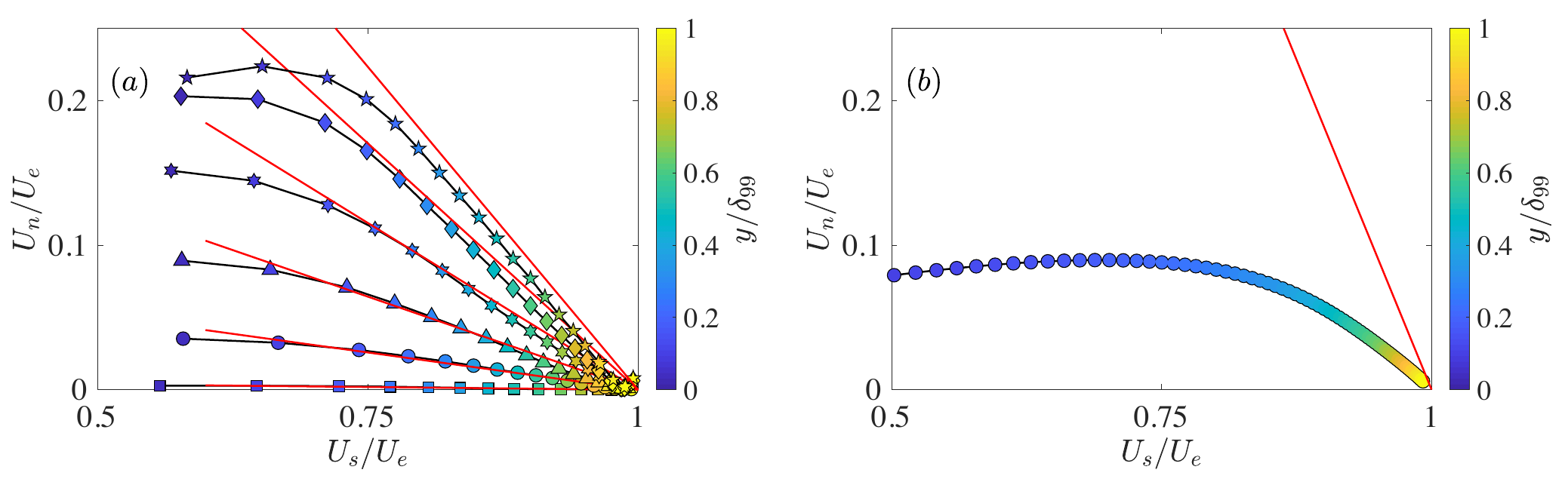}}

\caption{Johnston triangular plot ($a$) WMLES (EQWM) of the bent square duct: square, station 0; circle, station 4; triangle, station 6; cross, station 8; diamond, station 10; star, station 12. ($b$) DNS of the shear-driven 3DTBL from the transient channel flow at $Re_{\tau}=546$ at $t^+=192$ \citep{LozanoDuran2020}. Red straight line, the SWH formula Eq.~(\ref{eq:vorticity_local_swh}). Color bar denotes the wall distance normalized by the local boundary layer thickness.}
\label{fig:triangualr plots}
\end{figure}

In the present work, the mean three-dimensionality in the outer layer  is created by the inviscid skewing mechanism, where the streamwise vorticity is produced by reorientation of the spanwise vorticity.  A popular way of representing the crossflow so developed  is the 
``Johnston triangular plot'' \citep{Johnston1960}, which is the triangular plot of $U_s$ against $U_n$, where $U_s$ and $U_n$ are along and normal to the local freestream direction, respectively. In particular, the outer-layer mean velocity profile of the skew-induced 3DTBLs can be accurately approximated by  the Squire-Winter-Hawthorne (SWH) relation \citep{Squire1951,Hawthorne1951,Bradshaw1987},
\begin{equation}\label{eq:vorticity_local_swh}
     \frac{U_{n}}{U_{e}}=2\gamma_{e}\left(1-\frac{U_{s}}{U_{e}}\right).
\end{equation}
where $\gamma_{e}$ is the freestream turning along the streamline.
This is a special case of the vorticity transport equation in which viscous terms and Reynolds stresses are neglected.  The SWH relation  shows up as a straight line with a negative slope toward the right end of the triangular plot.  In Fig.~\ref{fig:triangualr plots}, we present the mean velocity for the current bent duct flow and a temporally developing shear-driven 3D channel flow  from \citet{LozanoDuran2020} in the triangular plot. 
It is observed that  the mean velocities
in the outer layer from  the duct flow satisfy the SWH formula well, whereas they deviate from the SWH relation in the shear-driven case,  as expected.  The slope in the SWH relation represents the freestream turning angle with respect to the upstream flow, and  the freestream slope in the triangular plot (Fig.~\ref{fig:triangualr plots}($a$)) therefore increases toward the downstream direction. 
\color{black}

\begin{figure}
\centerline{\includegraphics[width=1.0\linewidth]{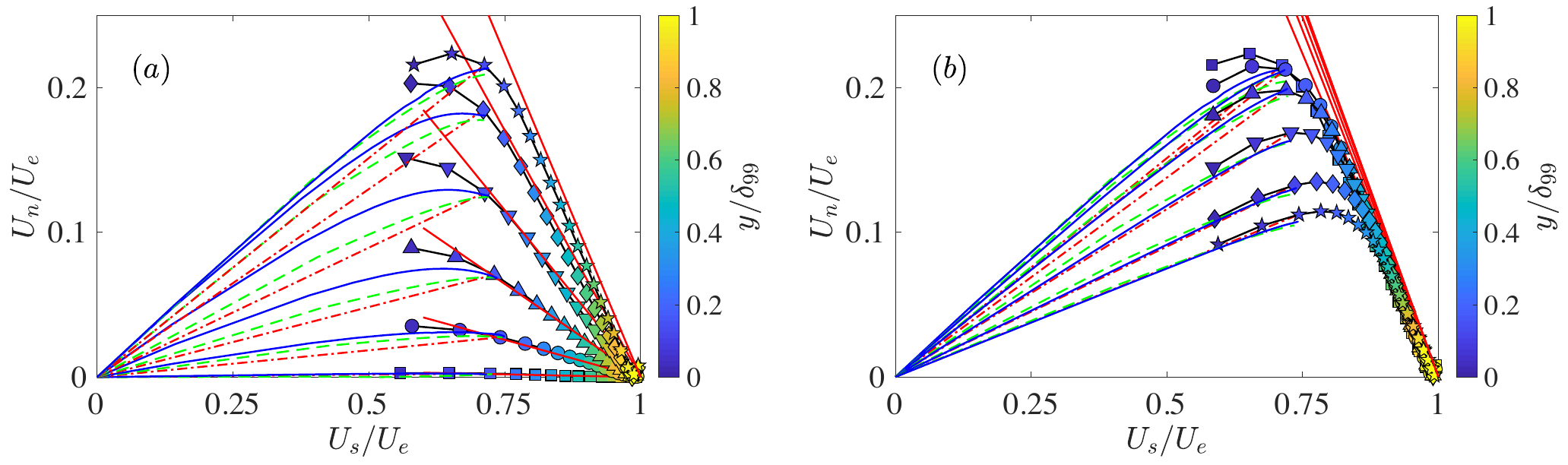}}
\caption{
Johnston triangular plot of the WM solutions. 
($a$) Crossflow developing stage: square, station 0; circle, station 4; triangle, station 6; cross, station 8; diamond, station 10; star, station 12. ($b$) Crossflow decaying stage: square, station 12; circle, station 14; triangle, station 16; cross, station 18; diamond, station 20; star, station 21. Red solid straight line, relation given by the SWH formula 
(Eq.~(\ref{eq:vorticity_local_swh})). Red dash-dotted line, EQWM; blue solid line, PDE NEQWM; green dashed line, integral NEQWM. Colorbar denotes the wall distance normalized by the local boundary layer thickness.
\color{black}}
\label{fig:triangualr plots wm}
\end{figure}

The wall model solutions (defined only close to the wall) 
are also visualized along with the outer LES solution  in Fig.~\ref{fig:triangualr plots wm}. It gives us a vivid picture of different wall models' capabilities to depict skewed mean-velocity profiles. In the triangular plot, the wall model solutions are from the origin to the 3rd cell LES solutions (coarse grid resolution). The EQWM, due to its unidirectional assumption, cannot describe skewed mean-velocity profiles, and it shows up as a straight line starting from the origin in the triangular plot. On the other hand, the PDE NEQWM and the integral NEQWM are able to represent skewed mean-velocity profiles. They show up as curved lines in the triangular plot, implying that the flow direction changes with the wall distance. During the crossflow developing stage (Fig.~\ref{fig:triangualr plots wm}($a$)), the difference between the two NEQWM solutions and the EQWM solution gradually grows. 
Notice that the PDE NEQWM is able to represent a richer variation of the slope (flow direction) along the wall-normal direction than the integral NEQWM. During the crossflow decaying stage (Fig.~\ref{fig:triangualr plots wm}($b$)), the difference among three wall model solutions gradually decreases. All three wall model solutions become almost unidirectional close to the end of the duct.

\color{black}
\subsection{\label{sec:level2}Quantification of the nonequilibrium contributions to the wall shear stress direction}
In this section, the nonequilibrium effects neglected in the EQWM are analyzed through the full RANS equations used in the nonequilibrium wall models. This analysis highlights the importance of accounting for the nonequilibrium effects in  wall modeling for accurate prediction of the surface flow direction, as well as the subtle difference in how these effects are incorporated in different wall models. 
Our analysis is based on the solutions of the PDE NEQWM and the integral NEQWM. Ideally, this analysis should be done in a priori sense utilizing the fully resolved flow fields, as attempted in \cite{Hickel2012}. This was deemed infeasible in the present case due to the high Reynolds number. 

Assuming incompressible flow, the time-averaged momentum equations in the PDE NEQWM can be recasted as 
\begin{eqnarray}\label{eq:rans_1}
&\frac{\partial}{\partial y}[(\nu+\nu_t)\frac{\partial \langle u\rangle }{\partial y}] = S_x
\end{eqnarray}
\begin{eqnarray}\label{eq:rans_2}
&\frac{\partial}{\partial y}[(\nu+\nu_t)\frac{\partial \langle w\rangle }{\partial y}] = S_z, 
\end{eqnarray}
where $S_x$ and $S_z$ are the nonequilibrium source terms comprising the  following terms ($S = A + P - D$),
\begin{eqnarray}
&\textrm{advection } A_x = \frac{\partial \langle u\rangle^2}{\partial x} + \frac{\partial \langle u\rangle \langle v\rangle }{\partial y} + \frac{\partial \langle u\rangle \langle w\rangle }{\partial z} +\frac{\partial \langle u'u'\rangle}{\partial x} + \frac{\partial \langle u'v'\rangle}{\partial y} + \frac{\partial \langle u'w'\rangle}{\partial z}\\
&\textrm{advection } A_z = \frac{\partial \langle w\rangle^2}{\partial z} + \frac{\partial \langle v\rangle \langle w\rangle }{\partial y} + \frac{\partial \langle u\rangle \langle w\rangle }{\partial x} +\frac{\partial \langle w'w'\rangle}{\partial z} + \frac{\partial \langle v'w'\rangle}{\partial y} + \frac{\partial \langle u'w'\rangle}{\partial x}\\
&\textrm{pressure gradient } P_x = \frac{1}{\rho}\frac{\partial \langle P\rangle}{\partial x}\\
&\textrm{pressure gradient } P_z = \frac{1}{\rho}\frac{\partial \langle P\rangle}{\partial z}\\
&\textrm{lateral diffusion } D_x = \frac{\partial}{\partial x}[(\nu+\nu_t)\frac{\partial \langle u\rangle }{\partial x}] + \frac{\partial}{\partial z}[(\nu+\nu_t)\frac{\partial \langle u\rangle }{\partial z}]\\
&\textrm{lateral diffusion } D_z = \frac{\partial}{\partial x}[(\nu+\nu_t)\frac{\partial \langle w\rangle }{\partial x}] + \frac{\partial}{\partial z}[(\nu+\nu_t)\frac{\partial \langle w\rangle }{\partial z}]
\end{eqnarray}

By integrating Eqns.~(\ref{eq:rans_1}) and (\ref{eq:rans_2}) twice,  the following expressions of the wall-shear stress for the PDE NEQWM are obtained 
\begin{eqnarray}\label{eq:tau_neqwm}
&\tau_{w,x} = \frac{U_{LES}-\int_{0}^{h_{wm}}\frac{\int_{0}^{y}S_xdy}{\nu+\nu_t}dy}{\int_{0}^{h_{wm}}\frac{1}{\nu+\nu_t}dy}, \\[8pt]
&\tau_{w,z} = \frac{W_{LES}-\int_{0}^{h_{wm}}\frac{\int_{0}^{y}S_zdy}{\nu+\nu_t}dy}{\int_{0}^{h_{wm}}\frac{1}{\nu+\nu_t}dy},
\end{eqnarray}
where $U_{LES}$ and $W_{LES}$ are the LES velocity components at the matching location. A similar expression can be found in \cite{Wang2002}. As shown in the previous sections, $U_{LES}$ and $W_{LES}$ are almost identical among the simulations with the different wall models. 
The wall shear stress direction, which is the quantity of interest 
exhibiting  the most significant difference among the three wall models, 
is then expressed as
\begin{equation}\label{eq:angle_neq}
    \frac{\tau_{w,z}}{\tau_{w,x}} = \frac{W_{LES}-I_z}{U_{LES}-I_x}, 
\end{equation}
where $I_z = \int_{0}^{h_{wm}}\frac{\int_{0}^{y}S_zdy}{\nu+\nu_t}dy$ and $I_x = \int_{0}^{h_{wm}}\frac{\int_{0}^{y}S_xdy}{\nu+\nu_t}dy$.
When all the nonequilibrium effects are neglected (i.e., letting $S_x=S_z=0$), 
this relation reduces to the wall-shear stress direction predicted by the EQWM which assumes unidirectional flow ($\frac{\tau_{w,z}}{\tau_{w,x}} = \frac{W_{LES}}{U_{LES}}$). 
The fidelity with which the constitutive terms of $I_x$ and $I_z$ are modeled is, therefore, crucial to the \color{black} performance of wall models in predicting the surface flow direction. 

To separate the nonequilibrium contributions from the flow direction predicted by the EQWM, we can first reorganize Eqn.~(\ref{eq:angle_neq}) as 
\begin{equation}\label{eq:angle_neq_recast}
    \frac{\tau_{w,z}}{\tau_{w,x}} = \frac{W_{LES}}{U_{LES}}\left(\frac{1-I_z/W_{LES}}{1-I_x/U_{LES}}\right).
\end{equation}
\color{black}

\begin{figure}
\centerline{\includegraphics[width=0.75\linewidth]{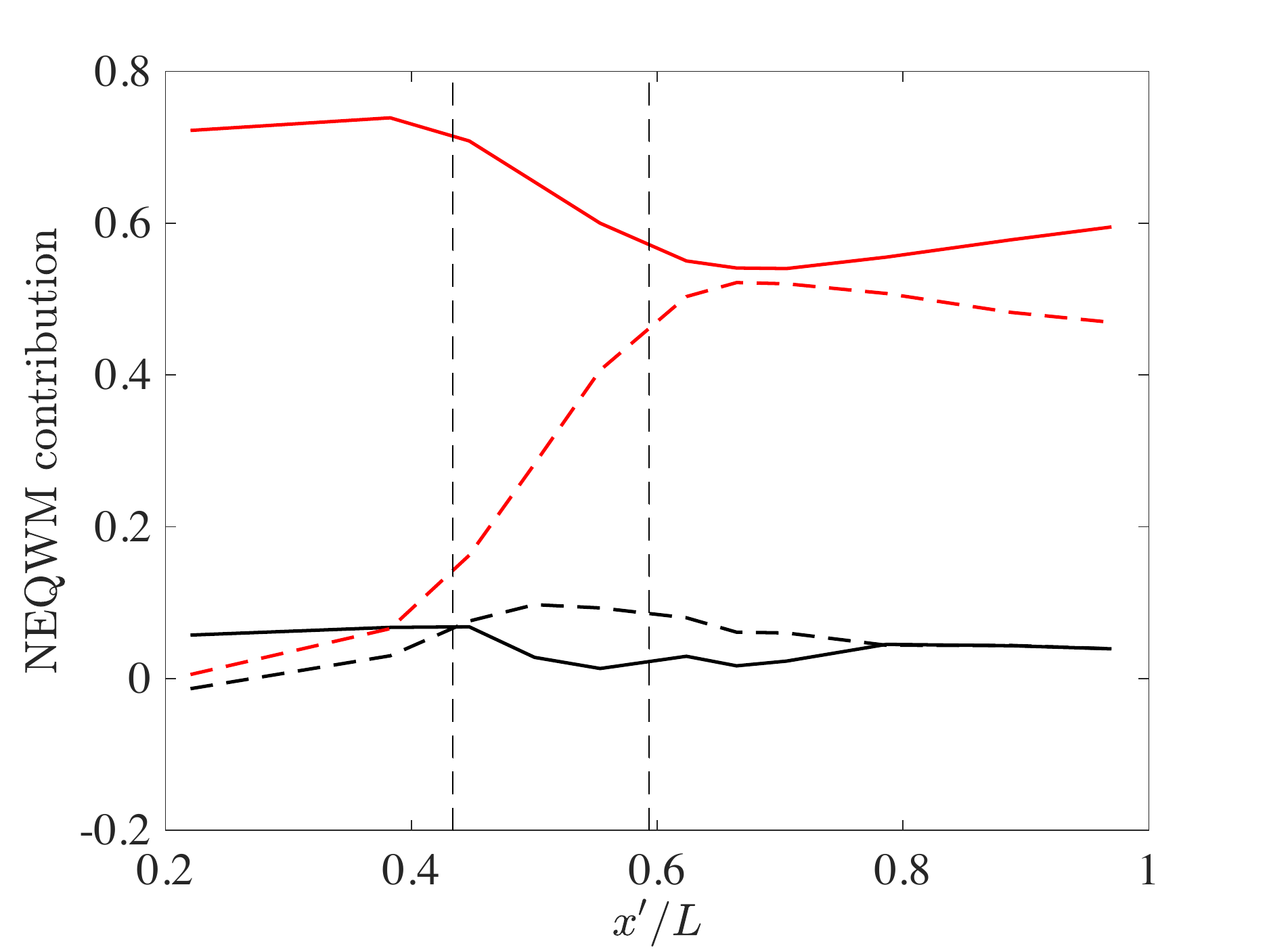}}
\caption{Centerline distribution of the nonequilibrium contributions. Red solid line, $U_{LES}/U_{ref}$; red dashed line, $W_{LES}/U_{ref}$; black solid line, $-I_x/U_{ref}$; black dashed line, $-I_z/U_{ref}$.}
\label{fig:neq_contribution}
\end{figure}

For the present flow, it is shown in Fig.~\ref{fig:neq_contribution} that $I_x$ is relatively small compared to $U_{LES}$, which permits the use of truncated Taylor series expansion of $\frac{1}{1-I_x/U_{LES}}$,

\begin{equation}\label{eq:angle_neq_taylor_denom}
    \frac{\tau_{w,z}}{\tau_{w,x}} = \frac{W_{LES}}{U_{LES}}\left(1-\frac{I_z}{W_{LES}}\right)\left[1+\frac{I_x}{U_{LES}}+O(\frac{I_x}{U_{LES}})^2\right], 
\end{equation}
\color{black}

Equation~(\ref{eq:angle_neq_taylor_denom}) can be further expanded as the following expression
\color{black}
\begin{equation}\label{eq:angle_taylor}
    \frac{\tau_{w,z}}{\tau_{w,x}} = \frac{W_{LES}}{U_{LES}}\left(1-\frac{I_z}{W_{LES}}+\frac{I_x}{U_{LES}}+\cdots\right). 
\end{equation}
\noindent The terms $I_x/U_{LES}$ and $I_z/W_{LES}$ can be viewed as representing the corrective contributions from the nonequilibrium effects to the surface flow direction predicted by the EQWM. In essence, Eqn.~(\ref{eq:angle_taylor}) enables the surface flow angle to be decomposed into distinct contributions originating from the equilibrium and the nonequilibrium terms. \color{black}
Although not shown here for brevity, this truncated relation was found to provide almost identical description as the actual wall flow direction 
($\tau_{w,z}/\tau_{w,x}$). 
Note that 
Eqn.~(\ref{eq:angle_taylor}) applies to the integral NEQWM as well. 
However, the terms $I_x$ and $I_z$ therein are largely modeled using  
the assumed velocity profile, in contrast to the PDE NEQWM where these terms are largely solved for, using the full RANS equations. \color{black}

\begin{figure}
\centerline{\includegraphics[width=0.75\linewidth]{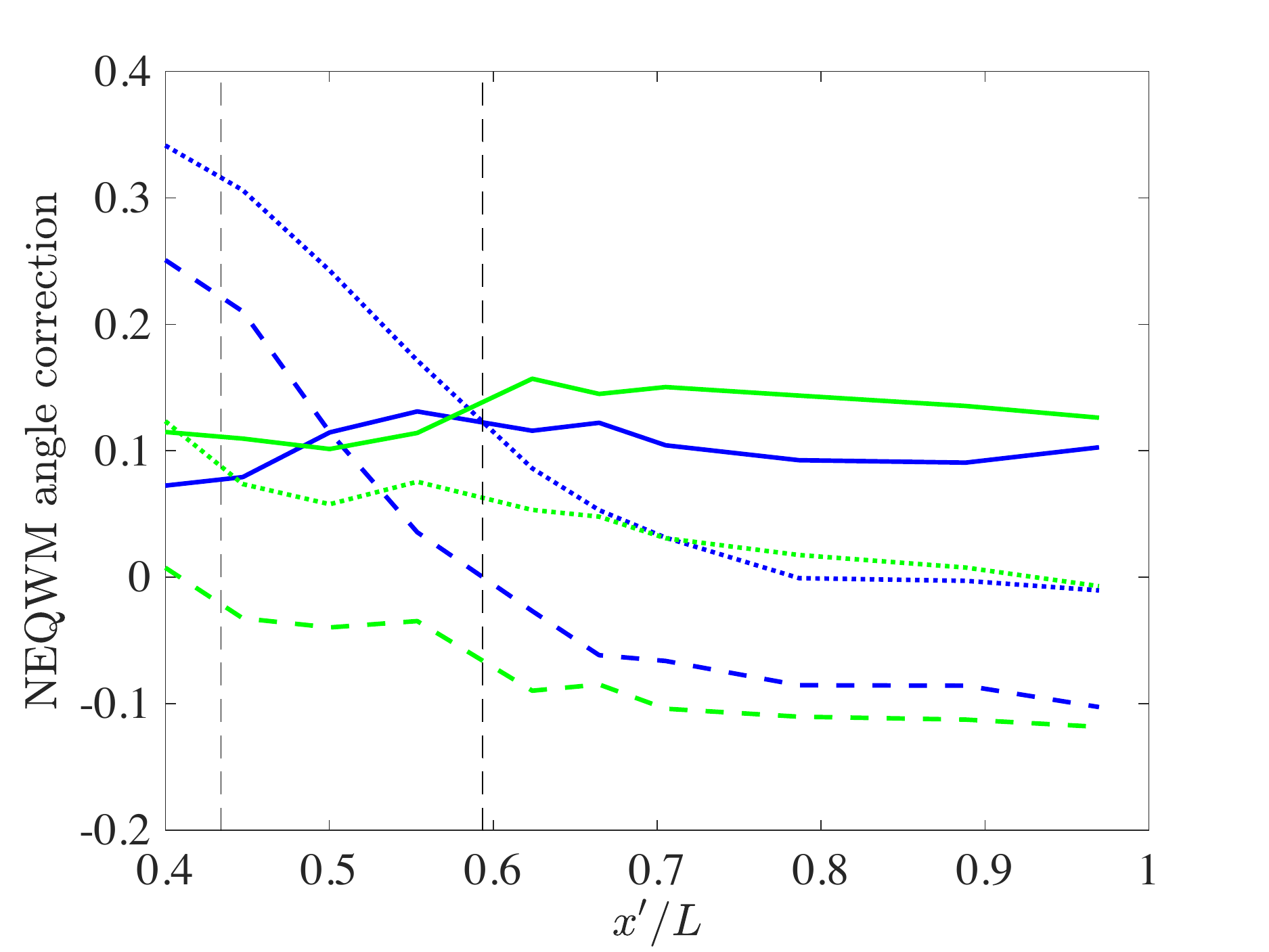}}
\caption{Centerline distribution of the nonequilibrium contribution to the flow direction. Solid line, $I_x/U_{LES}$; dashed line, $-I_z/W_{LES}$; dotted line, total nonequilibrium correction. Blue, NEQWM; green, integral NEQWM. Black vertical dashed lines, start and end of the bend region.}
\label{fig:neq_angle_correction}
\end{figure}

To further compare the wall models, we plot the two leading order terms of Eqn.~(\ref{eq:angle_taylor}) for the two nonequilibrium wall models in Fig.~\ref{fig:neq_angle_correction} (all nonequilibrium terms are zero for EQWM and they are not plotted.) 
Several interesting observations are made. 
First, the two nonequilibrium wall models show that the total nonequilibrium angle corrections are large at the beginning of the bend region, and that they gradually decrease to zero towards the end of the duct. That is, the nonequilibrium models properly sense the region where nonequilibrium effects are important, and attempt to model them therein. 
Second, 
the two nonequilibrium wall models produce comparable distributions of $I_x/U_{LES}$, implying that the axial ($x$) contents of the nonequilibrium effects are modeled almost identically by the two wall models. 
Third, the difference between the wall models appears to be  concentrated in $-I_z/W_{LES}$, i.e., in the way the models sense and model the cross-flow ($z$) component of the nonequilibrium effects. 
The integral NEQWM underpredicts $-I_z/W_{LES}$ throughout the duct compared to the PDE NEQWM. 
Furthermore, within the bend, the signs of this term are opposite in the two wall models.
This term is comprised of advection ($A_z$), pressure gradient ($P_z$), lateral diffusion ($D_z$), and $W_{LES}$. Among these terms, $P_z$ and $W_{LES}$ are imposed largely from the LES solution (which are seen to be identical from the simulations using the two wall models), and $D$ is seen to be negligible in its magnitude. Therefore, 
we conjecture that the difference of $-I_z/W_{LES}$ in the integral wall model originates largely from its assumed velocity profile, which is directly used in computing the advection term $A_{z}$\color{black}.  



\begin{figure}
\centerline{\includegraphics[width=1.0\linewidth]{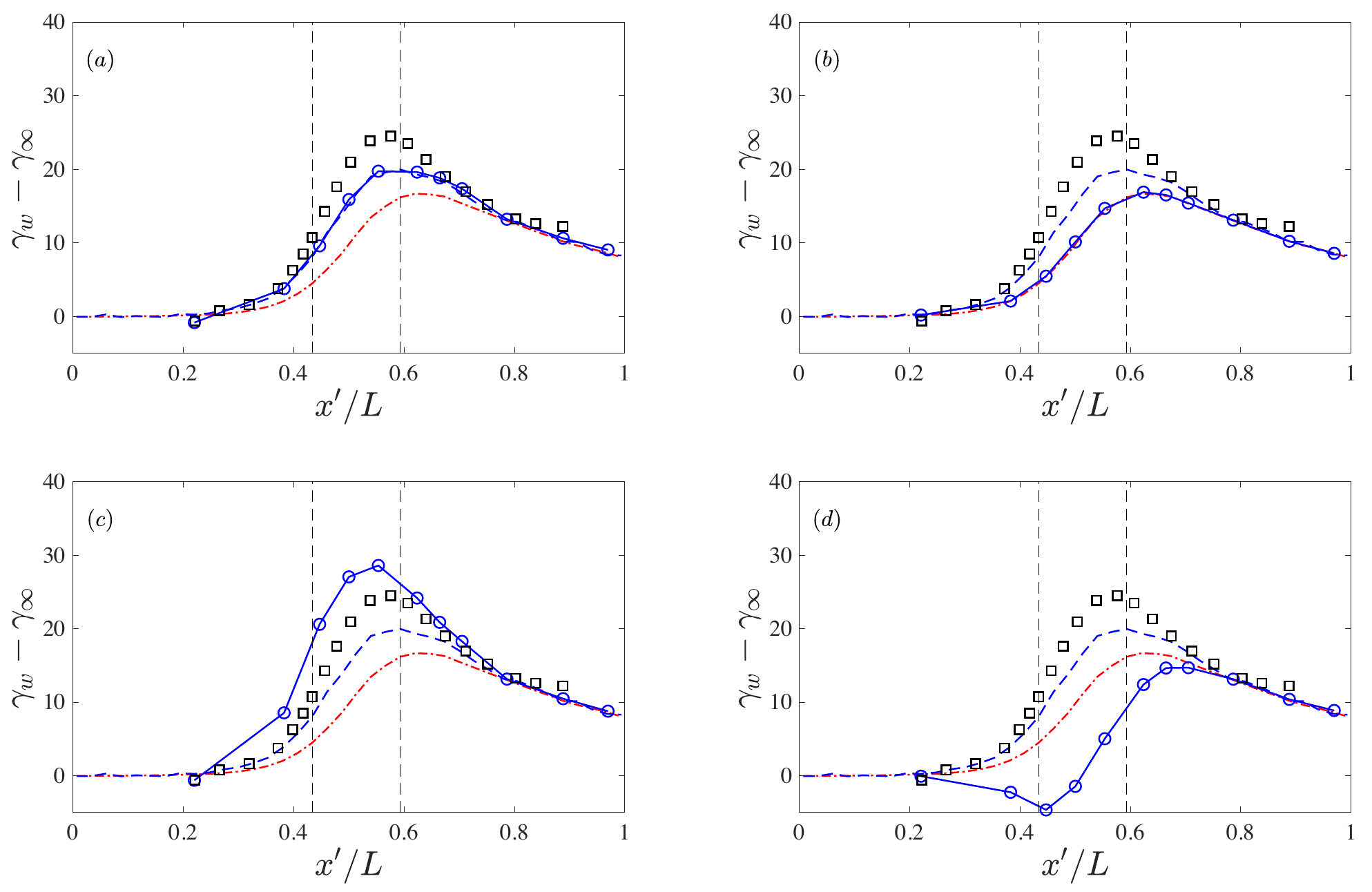}}
\caption{Centerline distribution of the surface flow turning angles with respect to the freestream ($\gamma_w$ is the wall shear stress direction, $\gamma_{\infty}$ is the freestream direction). ($a$) including all nonequilibrium effects, ($b$) including diffusion only, ($c$) including pressure gradient only, ($d$) including advection only. Red dash-dotted line, EQWM; blue dashed line, PDE NEQWM; blue solid line with circles, reconstruction with nonequilibrium contributions; black squares, experiment; black vertical dashed line, start and end of the bend region.}
\label{fig:angle_reconstruct}
\end{figure}

Furthermore, the individual contribution from different nonequilibrium effects are also analyzed. Starting from  Eqn.~(\ref{eq:angle_neq}) 
with $I_x = I_z =0$ (corresponding to the EQWM prediction), 
we examine the change in the surface flow turning angle by 
 systematically including the contributions from various nonequilibrium effects (advection, pressure gradient, and lateral diffusion) to 
 $I_x$ and $I_z$. Note that this is done in a post-processing manner using the solution of the PDE NEQWM.  
 The results are shown in Fig.~\ref{fig:angle_reconstruct}. When all the nonequilibrium effects are taken into account, the reconstructed surface flow turning-angle agrees with the PDE NEQWM prediction, as expected. The lateral  diffusion terms have negligible contributions to the surface flow turning angle. The pressure gradient and advection terms are significant within the bend, where the mean three-dimensionality develops. However, these terms appear to have a competing effect within the bend. \color{black} The pressure gradient tends to make the flow deviate more from the local freestream, while the advection tends to make the flow deviate less from the local freestream. 
 These two contributions largely cancel out each other, but a subtle balance between the two appears to be crucial in prediction of the surface flow direction. 

\color{black}
\section{\label{sec:conclusion}Conclusion}
In the present work, we have studied a spatially-developing pressure-driven 3DTBL over the floor of a square duct with a $30^{\circ}$ bend using WMLES. The major focus of this study is to contrast the performance of three commonly used wall models in a high Reynolds number pressure-driven 3DTBL. 

These models represent varying degrees of physical fidelity: the EQWM solves the simplified boundary-layer equations which neglect all nonequilibrium effects and assume the flow is unidirectional in the wall-modeled region; the PDE NEQWM solves unsteady 3D RANS equations which maintain much of the nonequilibrium effects; the integral NEQWM
represents  a compromise between the two models. Algebraic complexity is kept low thanks to the presumed velocity profile, while the nonequilibrium effect is represented by the linear perturbation to the logarithmic law of the wall.
\color{black}

The mean flow statistics and the Reynolds stresses are predicted reasonably well by the WMLES under the current grid resolutions, which are too coarse for the no-slip LES. The more comprehensive PDE NEQWM does show improvement against the integral NEQWM (which is in turn better than the EQWM) in predicting the direction of mean wall shear stress. 
The error in the local wall-shear force direction accumulates along the surface streamlines, leading to significant difference of the surface flow at the end of the duct. 
Budget analyses have been conducted to  elucidate precise mechanisms by which the three wall models produce different predictions of the wall-shear stress directions given almost identical inputs. For the present flow, the surface flow direction is shown to have separable contributions from the equilibrium part of the wall models and the integrated nonequilibrium effects (advection and pressure transports). It was shown that difference in how the cross-flow component of the nonequilibrium contribution is modeled leads to different behaviors of the models. Additionally, the pressure gradient and the advection are shown to have a competing effect  in deflecting the surface flow within the bend. Although these terms largely cancel out each other, neglecting any of them produces large errors in the surface flow direction, and 
a subtle balance between the two appears to be crucial in prediction of the surface flow. 

However, such difference in the wall shear stress direction 
predicted by the wall models appears to be not felt by the (outer) LES solution. The three wall models produce almost the same mean velocity and Reynolds stresses profiles. A possible explanation for this phenomenon comes from the nature of the pressure-driven 3DTBL. In the duct flow under consideration, the mean flow three-dimensionality in the outer layer is largely controlled by the ``inviscid skewing'' mechanism, which is not affected by the near wall viscous effects. 
In this class of 3DTBL, the outer-layer mean flow appears to be robustly set up by the inviscid effect, provided that the momentum drain by the wall is specified with reasonable accuracy only (in particular, its magnitude rather than the direction).

The characteristics of the 3DTBL are also analyzed. The anisotropy of turbulence 
along the wall-normal direction is investigated with the Lumley triangle. Compared to the 2DTBL, a large in-ward pointing sharp corner is presented in the Lumley triangle plot of 3DTBL in the downstream section. This large sharp corner represents a non-monotonic decrease of anisotopy for increasing wall distance.  
\color{black}
When the mean crossflow is generated by the inviscid effect (reorientation of spanwise vorticity), the relation between spanwise and streamwise velocity in the local freestream coordinate system in the outer part of the boundary layer shows as a straight line predicted by the SWH formula. In the present duct flow, the outer LES results show good agreement with the SWH formula, which shows that the``inviscid skewing'' mechanism is the major effect for generating mean three-dimensionality in the outer layer. 
The triangular plot of the inner wall-model solution reveals that 
the PDE NEQWM is most capable of representing the change in the flow direction along the wall-normal direction. The EQWM is most restrictive in this sense with its unidirectional flow assumption. The integral NEQWM is in between the two. 
\color{black}

\section{\label{sec:level1}Acknowledgement}
This research was sponsored by NASA's Transformational Tools and Technologies Project of the Transformative Aeronautics Concepts Program under the Aeronautics Research Mission Directorate (Grant 80NSSC18M0155), and by the Office of Naval Research (ONR) (Grant N000141712310). Computational resources supporting this work were provided by the NASA High-End Computing Program through the NASA Advanced Supercomputing Division at Ames Research Center.
\color{black}


\bibliographystyle{jfm}
\bibliography{jfm}

\begin{thebibliography}{67}
\expandafter\ifx\csname natexlab\endcsname\relax\def\natexlab#1{#1}\fi
\def\au#1{#1} \def\ed#1{#1} \def\yr#1{#1}\def\at#1{#1}\def\jt#1{\textit{#1}}
  \def\bt#1{#1}\def\bvol#1{\textbf{#1}} \def\vol#1{#1} \def\pg#1{#1}
  \def\publ#1{#1}\def\arxiv#1{#1}\def\org#1{#1}\def\st#1{\textit{#1}}

\bibitem[Bae {\em et~al.\/}(2019)Bae, Lozano-Dur{\'a}n, Bose \&
  Moin]{bae2019dynamic}
{\sc \au{Bae, H.~J.}, \au{Lozano-Dur{\'a}n, A.}, \au{Bose, S.~T.} \& \au{Moin,
  P.}} \yr{2019}  \at{Dynamic slip wall model for large-eddy simulation}.
  \jt{J. Fluid Mech.}  \bvol{859},  \pg{400--432}.

\bibitem[Balaras {\em et~al.\/}(1996)Balaras, Benoccis \&
  Piomelli]{Balaras1996}
{\sc \au{Balaras, E.}, \au{Benoccis, C.} \& \au{Piomelli, U.}} \yr{1996}
  \at{Two-layer approximate boundary conditions for large-eddy simulations}.
  \jt{\emph{AIAA J.}}  \bvol{34},  \pg{1111--1119}.

\bibitem[Bentaleb \& Leschziner(2013)]{Bentaleb2013}
{\sc \au{Bentaleb, Y.} \& \au{Leschziner, M.~A.}} \yr{2013}  \at{The structure
  of a three-dimensional boundary layer subjected to streamwise-varying
  spanwise-homogeneous pressure gradient}.  \jt{\emph{Int. J. Heat. Fluid.
  Fl.}}  \bvol{43},  \pg{109--119}.

\bibitem[Bissonnette \& Mellor(1974)]{Bissonnette1974}
{\sc \au{Bissonnette, L.~R.} \& \au{Mellor, G.~L.}} \yr{1974}  \at{Experiments
  on the behaviour of an axisymmetric turbulent boundary layer with a sudden
  circumferential strain}.  \jt{\emph{J. Fluid Mech.}}  \bvol{63},
  \pg{369--413}.

\bibitem[Bodart \& Larsson(2011)]{Bodart2011}
{\sc \au{Bodart, J.} \& \au{Larsson, J.}} \yr{2011}  \at{Wall-modeled large
  eddy simulation in complex geometries with application to high-lift devices}.
   \jt{\emph{Annual Research Briefs} in Center for Turbulence Research,
  Stanford University}  \pg{pp. 37--48}.

\bibitem[Bodart \& Larsson(2012)]{Bodart2012conference}
{\sc \au{Bodart, J.} \& \au{Larsson, J.}} \yr{2012}  \at{Wall-modeled large
  eddy simulation of the mcdonnell-douglas 30p/30n high-lift airfoil in
  near-stall conditions}.  \jt{\emph{30th AIAA Applied Aerodynamic Conference}}
   \pg{p. 3022}.

\bibitem[Bose \& Moin(2014)]{Bose2014}
{\sc \au{Bose, S.~T.} \& \au{Moin, P.}} \yr{2014}  \at{A dynamic slip boundary
  condition for wall-modeled large-eddy simulation}.  \jt{\emph{Phys. Fluids}}
  \bvol{26},  \pg{015104}.

\bibitem[Bose \& Park(2018)]{Park2018}
{\sc \au{Bose, S.~T.} \& \au{Park, G.~I.}} \yr{2018}  \at{Wall-modeled
  large-eddy simulation for complex turbulent flows}.  \jt{\emph{Ann. Rev.
  Fluid Mech.}}  \bvol{50},  \pg{535--561}.

\bibitem[Bradshaw(1987)]{Bradshaw1987}
{\sc \au{Bradshaw, P.}} \yr{1987}  \at{Turbulent secondary flows}.
  \jt{\emph{Ann. Rev. Fluid Mech.}}  \bvol{19},  \pg{53--74}.

\bibitem[Bradshaw \& Pontikos(1985)]{Bradshaw1985}
{\sc \au{Bradshaw, P.} \& \au{Pontikos, N.}} \yr{1985}  \at{Measurements in the
  turbulent boundary layer on an ‘infinite’ swept wing}.  \jt{\emph{J.
  Fluid Mech.}}  \bvol{159},  \pg{105--130}.

\bibitem[Bruns {\em et~al.\/}(1999)Bruns, Fernholz \& Monkewitz]{Bruns1999}
{\sc \au{Bruns, J.~M.}, \au{Fernholz, H.~H.} \& \au{Monkewitz, P.~A.}}
  \yr{1999}  \at{An experimental investigation of a three-dimensional turbulent
  boundary layer in an ‘s’-shaped duct}.  \jt{\emph{J. Fluid Mech.}}
  \bvol{393},  \pg{175--213}.

\bibitem[Cabot \& Moin(2000)]{Cabot2000}
{\sc \au{Cabot, W.} \& \au{Moin, P.}} \yr{2000}  \at{Approximate wall boundary
  conditions in large-eddy simulation of high reynolds number flow}.
  \jt{\emph{Flow, Turbul. Combust.}}  \bvol{63},  \pg{269--291}.

\bibitem[Chesnakas \& Simpson(1994)]{Chesnakas1994}
{\sc \au{Chesnakas, C.~J.} \& \au{Simpson, R.~L.}} \yr{1994}  \at{Full
  three-dimensional measurements of the cross-flow separation region of a 6:1
  prolate spheroid}.  \jt{\emph{Exp. Fluids}}  \bvol{17},  \pg{68--74}.

\bibitem[Cho {\em et~al.\/}(2021)Cho, Lozano-Dur\'an, Moin \& Park]{Cho2021}
{\sc \au{Cho, M.}, \au{Lozano-Dur\'an, A.}, \au{Moin, P.} \& \au{Park, G.~I.}}
  \yr{2021}  \at{Wall-modeled large-eddy simulation of turbulent boundary
  layers with mean-flow three-dimensionality}.  \jt{\emph{AIAA J.}}  \bvol{59},
   \pg{1707--1717}.

\bibitem[Choi \& Moin(2012)]{Choi2012}
{\sc \au{Choi, H.} \& \au{Moin, P.}} \yr{2012}  \at{Grid-point requirements for
  large eddy simulation: Chapman’s estimates revisited}.  \jt{\emph{Phys.
  Fluids}}  \bvol{24},  \pg{011702}.

\bibitem[Chung \& Pullin(2009)]{Chung2009}
{\sc \au{Chung, D.} \& \au{Pullin, D.~I.}} \yr{2009}  \at{Large-eddy simulation
  and wall modelling of turbulent channel flow}.  \jt{\emph{J. Fluid Mech.}}
  \bvol{631},  \pg{281--309}.

\bibitem[Coleman {\em et~al.\/}(2000)Coleman, Kim \& Spalart]{Coleman2000}
{\sc \au{Coleman, G.~N.}, \au{Kim, J.} \& \au{Spalart, P.~R.}} \yr{2000}  \at{A
  numerical study of strained three-dimensional wall-bounded turbulence}.
  \jt{\emph{J. Fluid Mech.}}  \bvol{416},  \pg{75--116}.

\bibitem[Coleman {\em et~al.\/}(2019)Coleman, Rumsey \& Spalart]{Coleman2019}
{\sc \au{Coleman, G.~N.}, \au{Rumsey, C.~L.} \& \au{Spalart, P.~R.}} \yr{2019}
  \at{Numerical study of a turbulent separation bubble with sweep}.
  \jt{\emph{J. Fluid Mech.}}  \bvol{880},  \pg{684--706}.

\bibitem[Deardorff(1970)]{Deardorff}
{\sc \au{Deardorff, J.~W.}} \yr{1970}  \at{The numerical study of three
  dimensional turbulent channel flow at large {Reynolds numbers}}.  \jt{J.
  Fluid Mech.}  \bvol{41},  \pg{453--480}.

\bibitem[Degraaff \& Eaton(2000)]{Degraaff2000}
{\sc \au{Degraaff, D.~B.} \& \au{Eaton, J.~K.}} \yr{2000}  \at{Reynolds-number
  scaling of the flat-plate turbulent boundary layer}.  \jt{\emph{J. Fluid
  Mech.}}  \bvol{422},  \pg{319--346}.

\bibitem[Driver \& Hebbar(1988)]{Driver1988}
{\sc \au{Driver, D.~M.} \& \au{Hebbar, S.~K.}} \yr{1988}  \at{Three-dimensional
  shear-driven boundary layer flow with streamwise adverse pressure gradient}.
  \jt{\emph{AIAA J.}}  \pg{p. 3661}.

\bibitem[Eitel-Amor {\em et~al.\/}(2014)Eitel-Amor, Orlu \&
  Schlatter]{Schlatter2014}
{\sc \au{Eitel-Amor, G.}, \au{Orlu, R.} \& \au{Schlatter, P.}} \yr{2014}
  \at{Simulation and validation of a spatially evolving turbulent boundary
  layer up to $re_{\theta}=8300$}.  \jt{\emph{Int. J. Heat. Fluid. Fl.}}
  \bvol{47},  \pg{57--69}.

\bibitem[Evans {\em et~al.\/}(2020)Evans, Lacy, Smith \& Rivers]{Evans2020}
{\sc \au{Evans, A.~N.}, \au{Lacy, D.~S.}, \au{Smith, I.} \& \au{Rivers, M.~B.}}
  \yr{2020}  \at{Test summary of the nasa high-lift common research model
  half-span at qinetiq 5-metre pressurized low-speed wind tunnel}.
  \jt{\emph{AIAA AVIATION Forum}} .

\bibitem[Fernholz \& Finley(1996)]{Fernholz1996}
{\sc \au{Fernholz, H.~H.} \& \au{Finley, P.~J.}} \yr{1996}  \at{The
  incompressible zero-pressure-gradient turbulent boundary layer: An assessment
  of the data}.  \jt{\emph{Prog. Aerospace Sci.}}  \bvol{32},  \pg{245--311}.

\bibitem[Flack \& Johnston(1994)]{Flack1994}
{\sc \au{Flack, K.} \& \au{Johnston, J.}} \yr{1994}  \at{Near-wall flow in a
  three-dimensional turbulent boundary layer on the endwall of a rectangular
  bend}.  \jt{\emph{32nd Aerospace Sciences Meeting and Exhibit}}  \pg{p. 405}.

\bibitem[Gao {\em et~al.\/}(2019)Gao, Zhang, Cheng \& Samtaney]{Gao2019}
{\sc \au{Gao, W.}, \au{Zhang, W.}, \au{Cheng, W.} \& \au{Samtaney, R.}}
  \yr{2019}  \at{Wall-modelled large-eddy simulation of turbulent flow past
  airfoils}.  \jt{\emph{J. Fluid Mech.}}  \bvol{873},  \pg{174--210}.

\bibitem[Goc {\em et~al.\/}(2021)Goc, Lehmkuhl, Park, Bose \&
  Moin]{goc2021large}
{\sc \au{Goc, K.~A.}, \au{Lehmkuhl, O.}, \au{Park, G.~I.}, \au{Bose, S.~T.} \&
  \au{Moin, P.}} \yr{2021}  \at{Large eddy simulation of aircraft at affordable
  cost: a milestone in computational fluid dynamics}.  \jt{Flow}  \bvol{1}.

\bibitem[Gr{\"o}tzbach(1987)]{grotzbach1987direct}
{\sc \au{Gr{\"o}tzbach, G.}} \yr{1987}  \at{Direct numerical and large eddy
  simulation of turbulent channel flows}.  \jt{Encyclopedia of fluid mechanics}
   \bvol{6},  \pg{1337--1391}.

\bibitem[Hawthorne(1951)]{Hawthorne1951}
{\sc \au{Hawthorne, W.~R.}} \yr{1951}  \at{Secondary circulation in fluid
  flow}.  \jt{\emph{Proc. R. Soc. Lond. A}}  \bvol{206},  \pg{374–387}.

\bibitem[Hayat \& Park(2021)]{Hayat2021}
{\sc \au{Hayat, I.} \& \au{Park, G.~I.}} \yr{2021}  \at{Numerical
  implementation of efficient grid-free integral wall models in
  unstructured-grid les solvers}.  \jt{arXiv preprint arXiv:2111.02542} .

\bibitem[Hickel {\em et~al.\/}(2012)Hickel, Touber, Bodart \&
  Larsson]{Hickel2012}
{\sc \au{Hickel, S.}, \au{Touber, E.}, \au{Bodart, J.} \& \au{Larsson, J.}}
  \yr{2012}  \at{A parametrized non- equilibrium wall-model for large-eddy
  simulations}.  \jt{\emph{Proceedings of the Summer Program 2012}, Center for
  Turbulence Research, Stanford Univ., Stanford, CA}  \pg{pp. 229--240}.

\bibitem[Hoyas \& Jimenez(2006)]{Hoyas2006}
{\sc \au{Hoyas, S.} \& \au{Jimenez, J.}} \yr{2006}  \at{Scaling of the velocity
  fluctuations in turbulent channels up to $re_{\tau}$ = 2003}.
  \jt{\emph{Phys. Fluids}}  \bvol{18},  \pg{011702}.

\bibitem[Johnston \& Flack(1996)]{Flackreview1996}
{\sc \au{Johnston, J.} \& \au{Flack, K.}} \yr{1996}  \at{Review—advances in
  three-dimensional turbulent boundary layers with emphasis on the wall-layer
  regions}.  \jt{\emph{J. Fluids Eng.}}  \pg{pp. 219--232}.

\bibitem[Johnston(1960)]{Johnston1960}
{\sc \au{Johnston, J.~P.}} \yr{1960}  \at{On the three-dimensional turbulent
  boundary layer generated by secondary flow}.  \jt{\emph{J. Basic Eng.}}
  \bvol{82},  \pg{233--246}.

\bibitem[Kawai \& Larsson(2012)]{Kawai2012}
{\sc \au{Kawai, S.} \& \au{Larsson, J.}} \yr{2012}  \at{Wall-modeling in large
  eddy simulation: Length scales, grid resolution, and accuracy}.
  \jt{\emph{Phys. Fluids}}  \bvol{24},  \pg{015105}.

\bibitem[Kawai \& Larsson(2013)]{Kawai2013}
{\sc \au{Kawai, S.} \& \au{Larsson, J.}} \yr{2013}  \at{Dynamic non-equilibrium
  wall-modeling for large eddy simulation at high reynolds numbers}.
  \jt{\emph{Phys. Fluids}}  \bvol{25},  \pg{015105}.

\bibitem[Khalighi {\em et~al.\/}(2011)Khalighi, Nichols, Ham, Lele \&
  Moin]{Khalighi2011}
{\sc \au{Khalighi, Y.}, \au{Nichols, J.~W.}, \au{Ham, F.}, \au{Lele, S.~K.} \&
  \au{Moin, P.}} \yr{2011}  \at{Unstructured large eddy simulation for
  prediction of noise issued from turbulent jets in various configurations}.
  \jt{\emph{17th AIAA/CEAS Aeroacoustics Conference 2011 (32nd AIAA
  Aeroacoustics Conference)}}  \bvol{101},  \pg{2886}.

\bibitem[Klein {\em et~al.\/}(2003)Klein, Sadiki \& Janicka]{Klein2003}
{\sc \au{Klein, M.}, \au{Sadiki, A.} \& \au{Janicka, J.}} \yr{2003}  \at{A
  digital filter based generation of inflow data for spatially developing
  direct numerical or large eddy simulations}.  \jt{\emph{J. Comput. Phys.}}
  \bvol{186},  \pg{652--665}.

\bibitem[Larsson(2021)]{Larsson2021}
{\sc \au{Larsson, J.}} \yr{2021}  \at{Simple inflow sponge for faster turbulent
  boundary-layer development}.  \jt{\emph{AIAA J.}}  \bvol{59},  \pg{1--3}.

\bibitem[Larsson {\em et~al.\/}(2016)Larsson, Kawai, Bodart \&
  Bermejo-Moreno]{Larsson2016}
{\sc \au{Larsson, J.}, \au{Kawai, S.}, \au{Bodart, J.} \& \au{Bermejo-Moreno,
  I.}} \yr{2016}  \at{Large eddy simulation with modeled wall-stress: recent
  progress and future directions}.  \jt{\emph{Mech. Eng. Rev.}}  \bvol{3},
  \pg{15--00418}.

\bibitem[Littell \& Eaton(1993)]{Littell1993}
{\sc \au{Littell, H.~S.} \& \au{Eaton, J.~K.}} \yr{1993}  \at{Experimental
  investigation of the three-dimensional boundary layer on a rotating disk}.
  \jt{\emph{Turbulent Shear Flows 8}}  \pg{pp. 403--414}.

\bibitem[Lohmann(1976)]{Lohmann1976}
{\sc \au{Lohmann, R.~P.}} \yr{1976}  \at{The response of a developed turbulent
  boundary layer to local transverse surface motion}.  \jt{\emph{J. Fluids
  Eng.}}  \bvol{98},  \pg{354--363}.

\bibitem[Lozano-Dur\'an {\em et~al.\/}(2021)Lozano-Dur\'an, Bose \&
  Moin]{Lozano2021}
{\sc \au{Lozano-Dur\'an, A.}, \au{Bose, S.~T.} \& \au{Moin, P.}} \yr{2021}
  \at{Performance of wall-modeled les with boundary-layer-conforming grids for
  external aerodynamics}.  \jt{\emph{AIAA J.}}  \pg{pp. 1--20}.

\bibitem[Lozano-Dur\'an {\em et~al.\/}(2020)Lozano-Dur\'an, Giometto, Park \&
  Moin]{LozanoDuran2020}
{\sc \au{Lozano-Dur\'an, A.}, \au{Giometto, M.~G.}, \au{Park, G.~I.} \&
  \au{Moin, P.}} \yr{2020}  \at{Non-equilibrium three-dimensional boundary
  layers at moderate reynolds numbers}.  \jt{\emph{J. Fluid Mech.}}
  \bvol{883},  \pg{A20}.

\bibitem[Owen {\em et~al.\/}(2020)Owen, Chrysokentis, Avila, Mira, Houzeaux,
  Borrell, Cajas \& Lehmkuhl]{Owen2020}
{\sc \au{Owen, H}, \au{Chrysokentis, G.}, \au{Avila, M.}, \au{Mira, D.},
  \au{Houzeaux, G.}, \au{Borrell, R.}, \au{Cajas, J.~C.} \& \au{Lehmkuhl, O.}}
  \yr{2020}  \at{Wall-modeled large-eddy simulation in a finite element
  framework}.  \jt{\emph{Int. J. Numer. Methods Fluids}}  \bvol{92},
  \pg{20--37}.

\bibitem[Park(2017)]{Park2017}
{\sc \au{Park, G.~I.}} \yr{2017}  \at{Wall-modeled large-eddy simulation of a
  high reynolds number separating and reattaching flow}.  \jt{\emph{AIAA J.}}
  \bvol{55},  \pg{3709–--3721}.

\bibitem[Park \& Moin(2014)]{Park2014}
{\sc \au{Park, G.~I.} \& \au{Moin, P.}} \yr{2014}  \at{An improved dynamic
  non-equilibrium wall-model for large eddy simulation}.  \jt{\emph{Phys.
  Fluids}}  \bvol{26},  \pg{015108}.

\bibitem[Park \& Moin(2016{\natexlab{{\em a\/}}})]{Park2016}
{\sc \au{Park, G.~I.} \& \au{Moin, P.}} \yr{2016{\natexlab{{\em a\/}}}}
  \at{Numerical aspects and implementation of a two-layer zonal wall model for
  les of compressible turbulent flows on unstructured meshes}.  \jt{\emph{J.
  Comput. Phys.}}  \bvol{305},  \pg{589--603}.

\bibitem[Park \& Moin(2016{\natexlab{{\em b\/}}})]{Park2016PRF}
{\sc \au{Park, G.~I.} \& \au{Moin, P.}} \yr{2016{\natexlab{{\em b\/}}}}
  \at{Space-time characteristics of wall-pressure and wall shear-stress
  fluctuations in wall-modeled large eddy simulation}.  \jt{\emph{Phys. Rev.
  Fluids}}  \bvol{1},  \pg{024404}.

\bibitem[Patel(1965)]{Patel1965}
{\sc \au{Patel, V.~C.}} \yr{1965}  \at{Calibration of the preston tube and
  limitations on its use in pressure gradients}.  \jt{\emph{J. Fluid Mech.}}
  \bvol{23},  \pg{185--208}.

\bibitem[Patterson {\em et~al.\/}(2021)Patterson, Balin \&
  Jansen]{Patterson2021}
{\sc \au{Patterson, J.}, \au{Balin, R.} \& \au{Jansen, K.}} \yr{2021}
  \at{Assessing and improving the accuracy of synthetic turbulence generation}.
   \jt{\emph{J. Fluid Mech.}}  \bvol{906},  \pg{R1}.

\bibitem[Piomelli \& Balaras(2002)]{Piomelli2002}
{\sc \au{Piomelli, U.} \& \au{Balaras, E.}} \yr{2002}  \at{Wall-layer models
  for large-eddy simulations}.  \jt{\emph{Ann. Rev. Fluid Mech.}}  \bvol{34},
  \pg{349--374}.

\bibitem[Piomelli {\em et~al.\/}(1989)Piomelli, Ferziger, Moin \& Kim]{shift}
{\sc \au{Piomelli, U.}, \au{Ferziger, J.}, \au{Moin, P.} \& \au{Kim, J.}}
  \yr{1989}  \at{New approximate boundary conditions for large eddy simulations
  of wall-bounded flows}.  \jt{Phys. Fluids A}  \bvol{1},  \pg{1061--1068}.

\bibitem[Poinsot \& Lele(1992)]{Poinsot1992}
{\sc \au{Poinsot, T.~J.} \& \au{Lele, S.~K.}} \yr{1992}  \at{Boundary
  conditions for direct simulations of compressible viscous flows}.
  \jt{\emph{J. Comput. Phys.}}  \bvol{101},  \pg{104--129}.

\bibitem[Pope(2000)]{Pope2000}
{\sc \au{Pope, S.~B.}} \yr{2000} {\em \emph{Turbulent flows}\/}.
  \publ{Cambridge University Press}.

\bibitem[Rumsey(2018)]{Rumsey2018}
{\sc \au{Rumsey, C.~L.}} \yr{2018}  \at{The nasa juncture flow test as a model
  for effective cfd/experimental collaboration}.  \jt{\emph{AIAA J.}}  \pg{p.
  3319}.

\bibitem[Sandberg(2012)]{Sandberg2012}
{\sc \au{Sandberg, R.}} \yr{2012}  \at{Numerical investigation of turbulent
  supersonic axisymmetric wakes}.  \jt{\emph{J. Fluid Mech.}}  \bvol{702},
  \pg{488--520}.

\bibitem[Schlatter {\em et~al.\/}(2010)Schlatter, Li, Brethouwer, Johansson \&
  Henningson]{Schlatter2010}
{\sc \au{Schlatter, P.}, \au{Li, Q.}, \au{Brethouwer, G.}, \au{Johansson,
  A.~V.} \& \au{Henningson, D.~S.}} \yr{2010}  \at{Simulations of spatially
  evolving turbulent boundary layers up to $re_{\theta}=4300$}.  \jt{\emph{Int.
  J. Heat. Fluid. Fl.}}  \bvol{31},  \pg{251--261}.

\bibitem[Schumann(1975)]{Schumann}
{\sc \au{Schumann, U.}} \yr{1975}  \at{Subgrid scale model for finite
  difference simulations of turbulent flows in plane channels and annuli}.
  \jt{J. Comput. Phys.}  \bvol{18},  \pg{376--404}.

\bibitem[Schwarz \& Bradshaw(1994)]{Schwarz1994}
{\sc \au{Schwarz, W.~R.} \& \au{Bradshaw, P.}} \yr{1994}  \at{Turbulence
  structural changes for a three-dimensional turbulent boundary layer in a 30
  $^{\circ}$ bend}.  \jt{\emph{J. Fluid Mech.}}  \bvol{272},  \pg{183--210}.

\bibitem[Sendstad(1992)]{Sendstad1992}
{\sc \au{Sendstad, O.}} \yr{1992}  \at{The near wall mechanics of
  three-dimensional turbulent boundary layers}.  \jt{\emph{Tech. Rep. Thermo
  Sci. Div. Mech. Eng. Stanford University}}  \bvol{TF 57}.

\bibitem[Spalart(1989)]{Spalart1989}
{\sc \au{Spalart, P.~R.}} \yr{1989}  \at{Theoretical and numerical study of a
  three-dimensional turbulent boundary layer}.  \jt{\emph{J. Fluid Mech.}}
  \bvol{205},  \pg{319--340}.

\bibitem[Squire \& Winter(1951)]{Squire1951}
{\sc \au{Squire, H.~B.} \& \au{Winter, K.~G.}} \yr{1951}  \at{The secondary
  flow in a cascade of airfoils in a nonuniform stream}.  \jt{\emph{J.
  Aeronaut. Sci}}  \bvol{18},  \pg{271}.

\bibitem[Vreman(2004)]{Vreman2004}
{\sc \au{Vreman, A.~W.}} \yr{2004}  \at{An eddy-viscosity subgrid-scale model
  for turbulent shear flow: Algebraic theory and applications}.
  \jt{\emph{Phys. Fluids}}  \bvol{16},  \pg{3670}.

\bibitem[Wang \& Moin(2002)]{Wang2002}
{\sc \au{Wang, M.} \& \au{Moin, P.}} \yr{2002}  \at{Dynamic wall modeling for
  large-eddy simulation of complex turbulent flows}.  \jt{\emph{Phys. Fluids}}
  \bvol{14},  \pg{2043}.

\bibitem[Yang {\em et~al.\/}(2017)Yang, Park \& Moin]{Yang2017}
{\sc \au{Yang, X. I.~A.}, \au{Park, G.~I.} \& \au{Moin, P.}} \yr{2017}
  \at{Log-layer mismatch and modeling of the fluctuating wall stress in
  wall-modeled large-eddy simulations}.  \jt{\emph{Phys. Rev. Fluids}}
  \bvol{2},  \pg{104601}.

\bibitem[Yang {\em et~al.\/}(2015)Yang, Sadique, Mittal \& Meneveau]{Yang2015}
{\sc \au{Yang, X. I.~A.}, \au{Sadique, J.}, \au{Mittal, R.} \& \au{Meneveau,
  C.}} \yr{2015}  \at{Integral wall model for large eddy simulations of
  wall-bounded turbulent flows}.  \jt{\emph{Phys. Fluids}}  \bvol{27},
  \pg{025112}.

\end{thebibliography}

\end{document}